\documentclass[default,iicol]{sn-jnl}

\jyear{2021}%
\usepackage{lettrine}
\usepackage{amssymb}
\usepackage{caption}
\usepackage{subcaption}
\usepackage{dblfloatfix}
\usepackage{cancel}
\usepackage{soul}
\usepackage{comment}
\usepackage{nameref}
\usepackage{bm}
\usepackage{breqn}
\usepackage[utf8]{inputenc}

\usepackage{csquotes}

\usepackage{wrapfig}

\let\OLDthebibliography\thebibliography
\renewcommand\thebibliography[1]{
  \OLDthebibliography{#1}
  \setlength{\parskip}{0pt}
  \setlength{\itemsep}{0pt plus 0.3ex}
}

\newcommand{\RomanNumeralCaps}[1]
    {\MakeUppercase{\romannumeral #1}}
    
\theoremstyle{thmstyletwo}%

\theoremstyle{thmstylethree}%

\begin{document}

\title[Article Title]{Design of an integrated hybrid plasmonic-photonic device for all-optical switching and reading of spintronic memory }


\author*[1,2]{\fnm{Hamed} \sur{Pezeshki}}\email{h.pezeshki@tue.nl}

\author[1]{\fnm{Pingzhi} \sur{Li}}
\author[1,2]{\fnm{Reinoud} \sur{Lavrijsen}}
\author[2]{\fnm{Martijn} \sur{Heck}}
\author[2]{\fnm{Erwin} \sur{Bente}}
\author[2]{\fnm{Jos} \sur{van der Tol}}

\author[1,2]{\fnm{Bert} \sur{Koopmans}}

\affil[1]{\orgdiv{Department of Applied Physics}, \orgname{Eindhoven University of Technology}, \orgaddress{ \city{Eindhoven}, \postcode{5612 AZ},  \country{Netherlands}}}

\affil[2]{\orgdiv{Eindhoven Hendrik Casimir Institute, Center for Photonic Integration}, \orgname{Eindhoven University of Technology}, \orgaddress{ \city{Eindhoven}, \postcode{5600 MB}, \country{Netherlands}}}


\abstract{
We introduce a novel integrated hybrid plasmonic-photonic device for all-optical switching and reading of nanoscale ferrimagnet bits. The racetrack memory made of synthetic ferrimagnetic material with a perpendicular magnetic anisotropy is coupled on to a photonic waveguide onto the indium phosphide membrane on silicon platform. The device which is composed of a double V-shaped gold plasmonic nanoantenna coupled with a photonic crystal cavity can enable switching and reading of the magnetization state in nanoscale magnetic bits by enhancing the absorbed energy density and polar magneto-optical Kerr effect (PMOKE) locally beyond the diffraction limit. Using a three-dimensional finite-difference time-domain method, we numerically show that our device can switch and read the magnetization state in targeted bits down to $\sim$100 nm in the presence of oppositely magnetized background regions in the racetrack with widths of 30 to 120 nm, clearly outperforming a bare photonic waveguide. Our hybrid device tackles the challenges of nonlinear absorption in the waveguide, weak PMOKE, and size mismatch between spintronics and integrated photonics. Thus, it provides missing link between the integrated photonics and nanoscale spintronics, expediting the development of ultrafast and energy efficient advanced on-chip applications.
}





\maketitle

\section*{Introduction}  \label{s-Intro}
The fields of integrated photonics and spintronics have been identified as two of the fastest growing directions for the new generation of solid-state application platforms. The integrated photonic platforms, such as silicon (Si)-on-insulator (SOI) \cite{bogaerts2005nanophotonic,bogaerts2012silicon} and indium phosphide (InP) membrane on silicon (IMOS) \cite{van2019inp,jiao2020indium}, enable high bandwidth and energy-efficient data transmission, which have achieved maturity over the years. In the meantime, spintronics promises the development of fast, energy-efficient, ultra-fast and -dense data memory devices \cite{Dieny:2020ub}. Much progress has already been made in the areas of magnetic random access memory (MRAM) \cite{Dieny:2020ub} with sizable commercialization. Meanwhile, a racetrack memory \cite{Parkin:2015aa}, in which trains of magnetic bits in magnetic nanoconduits are controlled employing cutting edge spintronic effects, has been proposed to further upgrade the low level cache in terms of speed, areal density and energy consumption \cite{BlaesingECT2018,Li:2022aa}.  
Logically, it can be envisioned that a further gain in the speed and energy efficiency can be achieved by merging photonics and spintronics domains into a single platform \cite{Kimel_AOS_Review2019}. Nevertheless, such a hybrid integration is not yet available since the communication between the two domains has to be controlled via the electronics acting as an intermediate domain, which creates power/time/area overhead, compromising the potential advantage of the hybrid integration. A possible way to mitigate this issue is the direct access (switching and reading) of spintronic memory by photonics.\par
 
So far, the all-optical reading (AOR) of a magnetic material exploiting the polar magneto-optical (MO) Kerr effect (PMOKE), i.e. the magnetization-dependent variation in the polarization state of light reflected off the surface of a magnetic material, has been used as a standard way for the reading of a standalone magnetic bit in the absence of oppositely magnetized neighboring bits \cite{maccaferri2016anisotropic,pineider2018nanomaterials,demirer2021magneto,demirer2022integrated,hamedAOR}. Recently, all-optical switching (AOS) \cite{Ostler:2012aa,Radu:2011aa,Stanciu:2007aa,Lalieu:2017aa} was shown to be a next step forward for optical access of the magnetic information from the fast advancing spintronic memory platforms \cite{Dieny:2020ub,Kimel_AOS_Review2019}. Recent advancements in AOS were embodied in the discovery of single pulse AOS in more process compatible synthetic ferrimagnets, such as cobalt (Co)/gadolinium (Gd) \cite{Lalieu:2017aa} and [Co/terbium (Tb)]$_m$ \cite{Aviles-Felix:2019aa,Aviles-Felix:2020aa} multilayered material platforms, having $m$ repetition(s). This approach provides a low switching energy \cite{Li:2021wr} and is compatible with integration to state-of-the-art spintronic building blocks \cite{Luding:2022ur, Wang:2020ab,Aviles-Felix:2020aa,Sobolewska:2020aa,Chen:2017aa}. 

Nevertheless, coupling light energy directly to spintronic devices comes with an obvious drawback of inefficient interaction due to the large size mismatch between the waveguide mode and the size of spintronic devices \cite{Dieny:2020ub}, as a result of the diffraction limit. Such an issue not only offsets the energy efficiency of AOS but also imposes challenges on the scalability of such a hybrid integration. Moreover, large transient power needed for AOS can incur additional losses in the waveguide due to nonlinear absorption \cite{Sobolewska:2020aa}. Similar issues also persist for AOR as weak coupling limits the MO interaction, resulting in a small PMOKE, even at large footprints. Therefore, a photonic design to focus light beyond the diffraction limit, while ensuring efficient energy transfer, is required. Here, we propose the concept of a hybrid plasmonic nanoantenna (PNA) and photonic crystal (PhC) cavity to curb the mentioned challenges. \par
 
Noble metal nanostructures such as gold PNAs can strengthen the light-matter interaction beyond the diffraction limit by localizing and enhancing incident electromagnetic energy in nanoscale spots at the metal interface using localized surface plasmon resonance (LSPR) \cite{hutter2004exploitation,gramotnev2010plasmonics}. The magnified light-matter interaction empowers PNAs to enhance the scattering and absorption cross sections of coupled nanoparticles, improving their effective polarizability and absorbed energy density. 
Such characteristics have turned PNAs into key components in designing miniaturized photonic devices in a variety of applications ranging from telecom \cite{pezeshki2021ultra,Pezeshki:22} to biosensing \cite{mejia2018plasmonic,pezeshki2021lab}. Therefore, it is expected that PNAs can likewise play a key role in enhancing the MO activity at the nanoscale and consequently addressing the scalability issue. \par

Despite large enhancement of the electric field provided by a PNA, the interaction cross section is still limited due to its small feature size. We propose the use of a PhC cavity \cite{zain2007tapered,pezeshki2015design,reniers2019characterization} to further improve the interaction cross section. A PhC cavity, basically a Bragg grating cavity based on periodic dielectric structures satisfying the Bragg condition \cite{yablonovitch1994photonic,joannopoulos1997photonic}, provides high spatial light confinement in a diffraction-limited cavity. The confined light in the cavity enhances the effective light-matter interaction cross section, making PhC cavities attractive for various applications such as lasing and optical switching \cite{ellis2011ultralow,pezeshki2013all,pezeshki2016design}. \par

With the above notions, we thus introduce the design of a hybrid PNA-PhC device coupled with a magnetic racetrack \cite{Parkin:2008aa,blasing2020magnetic,Li:2022aa} to bring all the advantages of both concepts in one place for performing enhanced AOS and PMOKE-based AOR of ferrimagnetic bits with perpendicular magnetic anisotropy (PMA) on a photonic integrated platform (Fig. \ref{fig 1}). Such a compact device allows for a direct energy-efficient transport of information from the spintronic domain to the photonic domain without need for intermediate electronics for signal conversion. Our hybrid device, which consists of a double V-shaped gold PNA coupled to a PhC cavity, can provide the threshold light fluence for AOS of Co/Gd magnetic films \cite{li2021ultra,Lalieu:2017aa,Li:2022aa} and improve the inherently weak PMOKE for AOR by efficiently focusing light onto a nanoscale spot in a magnetic racetrack using LSPR. Note that our Co/Gd based magnetic racetrack encodes information as domains with up and down magnetization states in a magnetic racetrack. These domains can be coherently moved at high velocity ($>$ 1 km/s) \cite{Li:2022aa} by the spin-orbit torque \cite{Miron:2011aa} and the Dzialoshiinskii-Moriya interaction \cite{Ryu:2013aa,BlaesingECT2018} driven domain wall motion, exerted by electrical current \cite{blasing2020magnetic}. We stress that our choice for PMA thin films is motivated by its leading role in state-of-the-art spintronics.\par

Based on a three dimensional finite-difference time-domain (3D FDTD) method \cite{fdtd},
we show that using our hybrid device, for an incident optical pulse energy of 0.6 pJ, can provide a switching fluence of 0.5 mJ/cm\textsuperscript{2} for switching magnetic bits with sizes of down to $\sim$100 nm in a racetrack with a width of 120 nm. Reducing the racetrack's width down to 30 nm, more enhanced light-matter interaction due to the stronger interaction between the PNA elements, further increases the absorbed energy density in the racetrack. Consequently, the switching of the magnetization in magnetic bits down to $\sim$60 nm can be realized for an ultralow incident pulse energy of 0.15 pJ. Moreover, the enhanced MO interaction offered by the hybrid device can enable the optical detection of magnetization in the targeted bits down to $\sim$100 nm in a racetrack with widths of 30 to 120 nm, regardless of the magnetization in the rest of the racetrack, with a relative contrast of 0.6\% for a (bit size $\times$ racetrack width) $\sim$ 200 $\times$ 100 nm\textsuperscript{2} magnetic bit. The proposed device is studied in the IMOS platform. However, our model utilizes generic concepts, which can be implemented in other popular photonic platforms such as SOI \cite{bogaerts2005nanophotonic}, silicon nitride \cite{subramanian2013low} and so on. We believe this device can be a significant step toward addressing the fundamental issue of hybrid spintronic-photonic integration imposed by the light diffraction limit and inefficient coupling, allowing the creation of new generation on-chip and inter-chip applications.\par

\section*{Design considerations} \label{s-Results}

         \begin{figure*}[hb!]
             \centering
                \begin{subfigure}[ht!]{0.8\textwidth}
                \centering
                \hspace*{-1.2cm} 
                 \includegraphics[scale = 0.5]{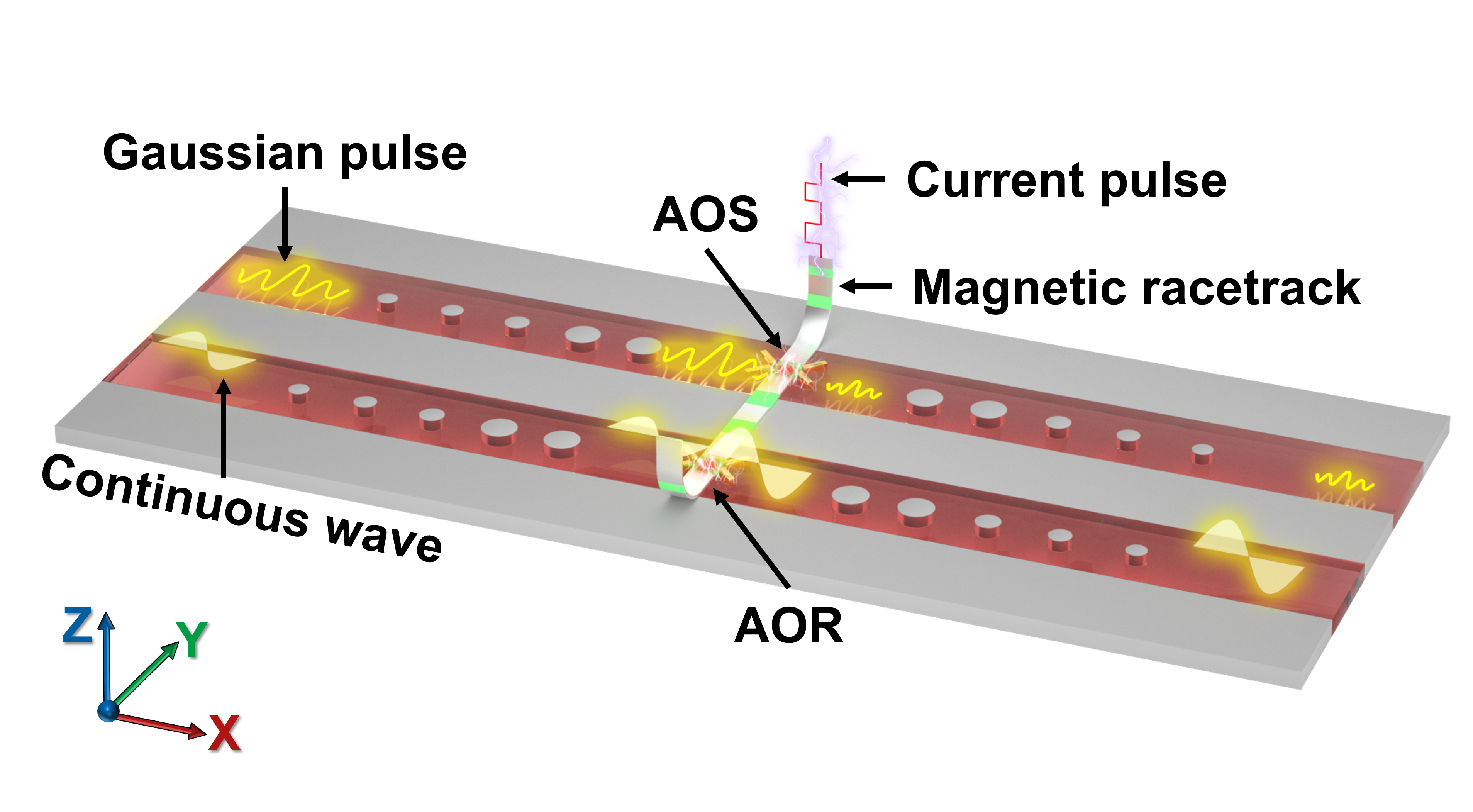}
                 \vspace{-3\baselineskip}
                 \caption{}
                   \label{fig 1a}
                 \end{subfigure}
                 \begin{subfigure}[ht!]{0.4\textwidth}
                \centering
                 \hspace*{-1.8cm} 
                 \vspace*{1.5cm}
                 \includegraphics[scale = 0.35]{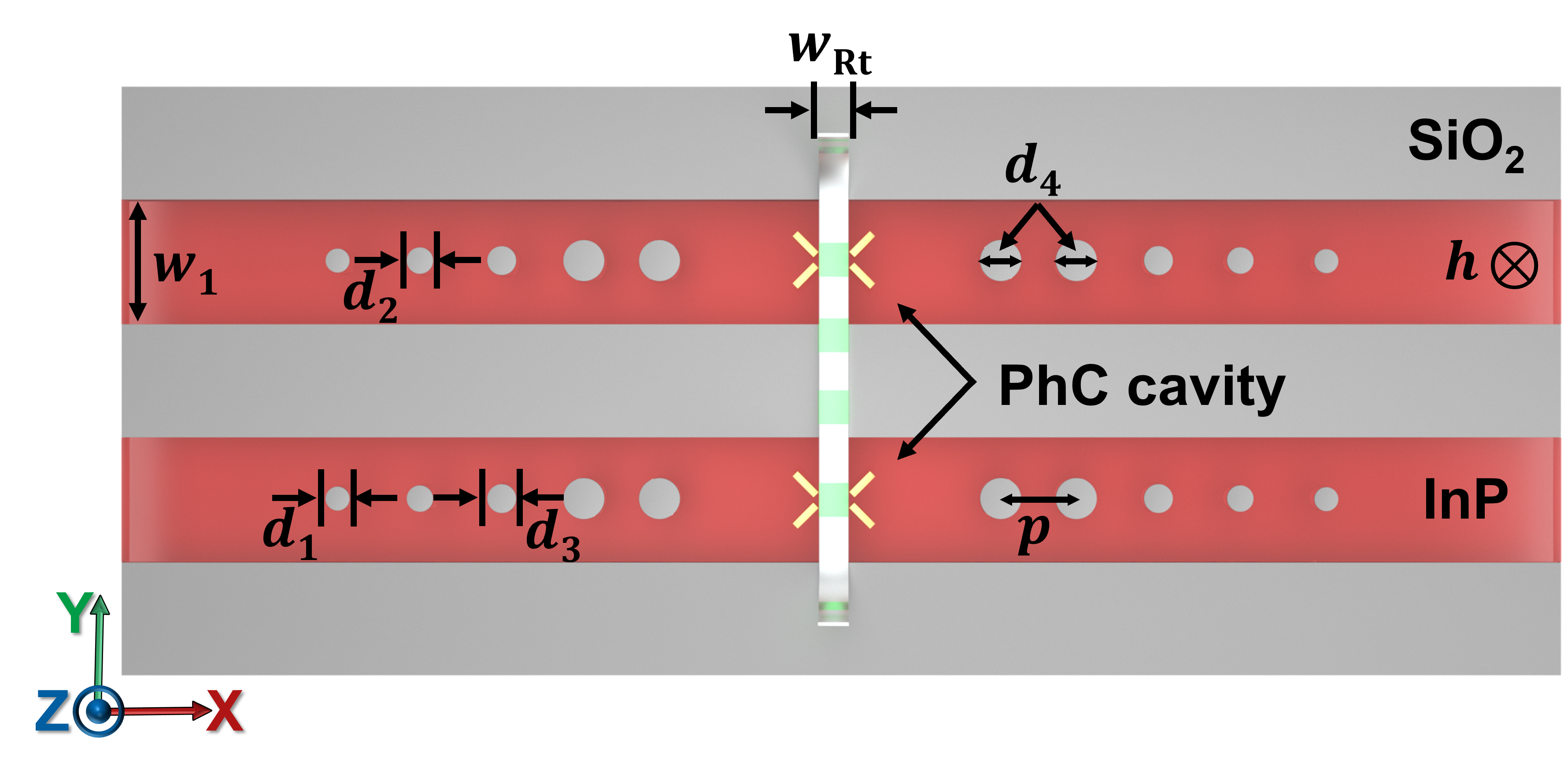}
                 \vspace{-5\baselineskip}
                \caption{\hspace*{-0.4cm}}
                 \label{fig 1b}
                \end{subfigure}
                \begin{subfigure}[ht!]{0.4\textwidth}
                \centering
                 \hspace*{1.4cm} 
                 \vspace*{1.5cm}
                 \includegraphics[scale = 0.35]{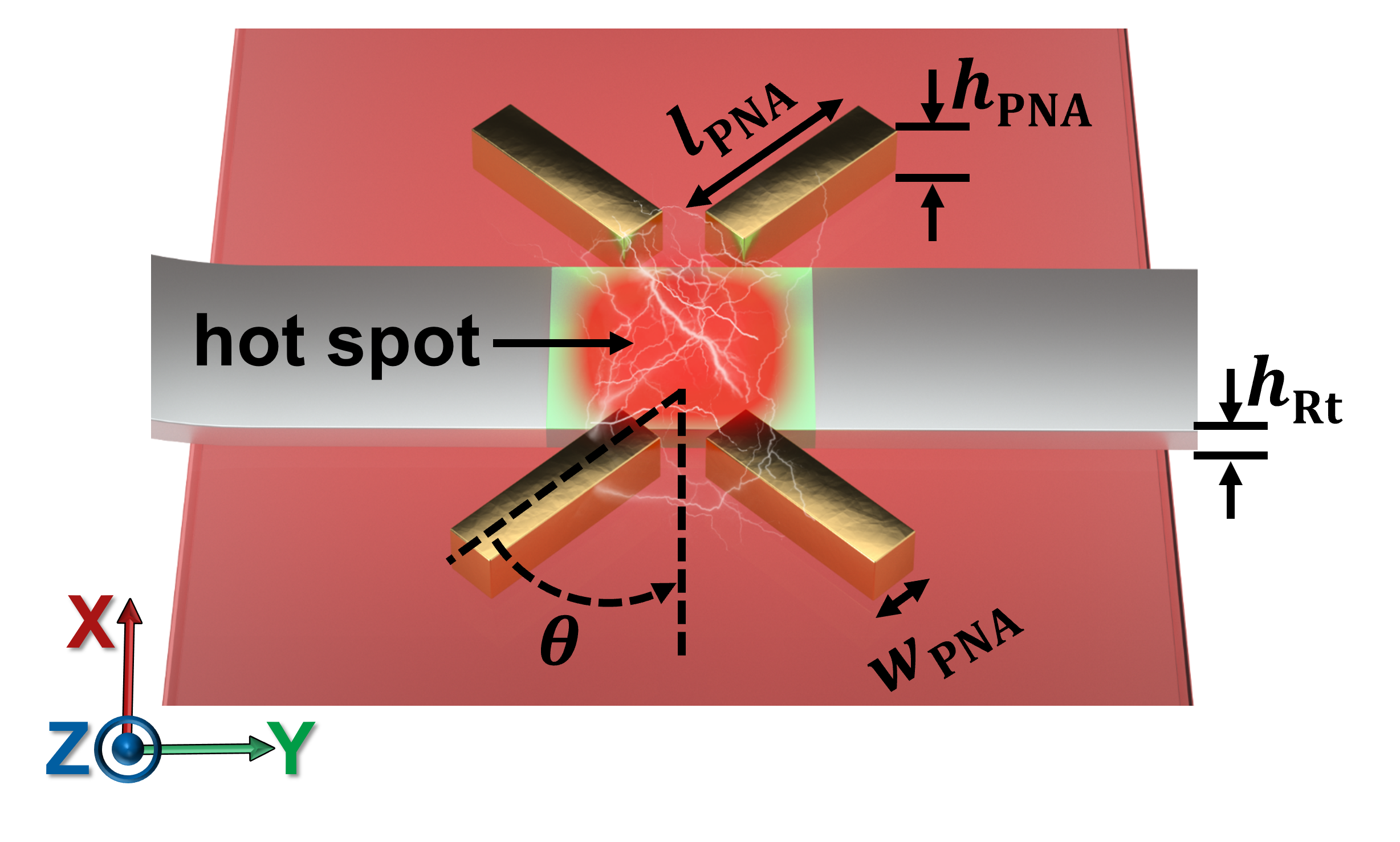}
                 \vspace{-4\baselineskip}
                  \caption{\hspace*{-3.2cm}}
                  \label{fig 1c}
                 \end{subfigure}
                  \centering
                \caption{Conceptual illustration of the hybrid all-optical switching/reading (AOS/AOR) device. (a) Perspective view showing how the proposed device functions. (b) Top view of the platform presenting design parameters of the photonic crystal (PhC) cavity, $d$\textsubscript{1-4} = 100, 110, 120, and 170 nm are the holes diameters with a pitch, $p$, of 370 nm. The waveguides have a width and a height of $w$\textsubscript{1} = 570 nm and $h$ = 280 nm, while the magnetic racetrack width, $w$\textsubscript{Rt}, is 120 nm. (c) Magnified view of the plasmonic nanoantenna (PNA) and the racetrack, where the length, height and width of the PNA elements are $l$\textsubscript{PNA} = 120 nm, $h$\textsubscript{PNA} = $w$\textsubscript{PNA} = 30 nm, and they are oriented at an angle of $\theta$ = 45$^\circ$ with reference to the waveguide direction. The racetrack height, $h$\textsubscript{Rt}, is 10 nm.}
                \label{fig 1}
                \end{figure*}

Within the scope of this work, we explore the MO interaction between the light in an InP waveguide in a standard IMOS platform and a magnetic racetrack as a top cladding of the waveguide coupled to a hybrid PNA-PhC. The platform concept is schematically depicted in Fig. \ref{fig 1}. It consists of two parallel InP waveguides with silica (SiO$_2$) side walls, where a racetrack is coupled orthogonally on top of the waveguides. For the occurrence of single-pulse AOS, a combination of a (3d) transition metal ferromagnet and a (4f) rare-earth ferromagnet is essential \cite{Mentink:2012aa,Ostler:2012aa,Kimel_AOS_Review2019}. For this reason, the racetrack consists of a multilayer stack of (from bottom to top) 4 nm heavy metal seed layer, a 2 nm ferromagnetic Co layer, and a 4 nm Gd layer (see \nameref{s-method}). A Gaussian pulse (for AOS) and a continuous wave (for AOR) both with a free space wavelength of 1.55 $\mu$m (in the telecom C-band) are coupled into the transverse electric (TE\textsubscript{0}) mode of both waveguides. The propagating light mode in the waveguide couples to the PhC cavity, where the light mode is spatially confined and enhanced at the center of the cavity. The spatially enhanced mode then evanescently excites the PNA by which light gets further focused and enhanced in a nano-spot at the central gap of the PNA. The concentrated and enhanced light energy interacts with the racetrack locally and a domain at the nano-spot is thermally switched \cite{Lalieu:2017aa,Mentink:2012aa,Beens:2019aa}. By injecting current pulses through the racetrack (see Fig. \ref{fig 1a}), the switched ferrimagnetic domains at the nano-spot of the racetrack on top of the waveguide is moved along the racetrack toward the bottom waveguide, where they can be read-out by the enhanced PMOKE in the nano-spot at the central gap of the PNA. Such a monolithic integration scheme gives room to the fabrication of spintronic building blocks directly in the existing platform, interfacing to both integrated photonics and complementary metal–oxide–semiconductor (CMOS) without requirements for a complex inter-chip integration.\par

The design parameters are indicated in Figs. \ref{fig 1b} and \ref{fig 1c}. The width and height of the racetrack are $w$\textsubscript{Rt} = 120 nm and $h$\textsubscript{Rt} = 10 nm, the width and height of the waveguides are $w$\textsubscript{1} = 570 nm and $h$ = 280 nm (see Supplementary note 1 for the design principle). The PhC cavity is designed by removing the central three holes from the PhC structure to create a resonant cavity at a wavelength of 1.55 $\mu$m (see Supplementary note 2). The holes toward the end of the PhC cavity are tapered down to lower the back-reflection by providing an impedance match between the waveguide mode and cavity mode. The pitch, $p$, is 370 nm and the holes diameters are $d$\textsubscript{1-4} = 100, 110, 120, and 170 nm, respectively. The PNA which is composed of two coupled V-shaped gold nanoantennas has a length, width and height of $l$\textsubscript{PNA} = 120 nm, $w$\textsubscript{PNA} = $h$\textsubscript{PNA} = 30 nm, and the PNA elements are oriented at an angle of $\theta$ = 45$^\circ$ with reference to the waveguide direction. The PNA is designed to have a resonance peak at 1.55 $\mu$m to optimize the light-matter interaction (see Supplementary note 3 for the design principle).\par

Next, in order to validate the above mentioned design in the coming sections, we will discuss the considerations regarding the power/energy limitation in a semiconductor waveguide, taking the example of an InP waveguide, as imposed by the known nonlinear effects. \par

\vline


\noindent \textbf{Energy considerations for all-optical switching.} AOS is considered as the least dissipative and the fastest way for magnetization switching \cite{Kimel_AOS_Review2019}. The AOS is accomplished by ultrafast excitation (heating) of free electrons by a light pulse and subsequent angular momentum transfer between two sub-lattices (Co and Gd in our case) \cite{Mentink:2012aa,Beens:2019aa,Gerlach:2017aa, Kimel_AOS_Review2019}. Nevertheless, the necessity of heating the metal non-adiabatically up to the Curie temperature point within a pulse duration \cite{Mentink:2012aa,Radu:2011aa,Ostler:2012aa} requires short pulses with a significant peak power. As a consequence, nonlinear losses, particularly, two photon absorption (TPA) and free-carrier absorption (FCA) \cite{Latkowski:2015uw,Thourhout:2001vm,Gonzalez:2009ut}, in a semiconductor photonic waveguide (e.g. made of Si and InP) might be incurred. Observations have been made previously both in IMOS \cite{Latkowski:2015uw} and SOI \cite{Sobolewska:2020aa} platforms, both showing significant nonlinear losses as the peak power of the pulses surpasses 0.1 W.\par

Previous works on AOS \cite{Lalieu:2017aa,Li:2021wr,Gorchon:2016aa,Li2022He,Ostler:2012aa} have demonstrated that an absorbed fluence threshold of around 0.5 mJ/cm\textsuperscript{2} is needed for switching a domain in a Co/Gd bilayer spintronic material stack, which corresponds to an energy deposition of 12 fJ to a waveguide's cladding with an area of 50 $\times$ 50 nm\textsuperscript{2} for a pulse shorter than 500 fs, which accounts for a peak power density of at least 1 GW/cm\textsuperscript{2}, which is close to the photon density inducing nonlinear losses \cite{Saleh:2007aa}. Due to the finite coupling efficiency between light in the waveguide and spintronic cladding as well as a limited interaction cross section, the incident peak power density essential for switching is expected to be even larger. Fortunately, earlier works \cite{Gorchon:2016aa,Wei:2021ui} showed successful AOS with a much longer $\sim$ 10 ps pulse, where the threshold energy for AOS has not been increased by more than a factor of two. Therefore, to get an idea of the power/energy limitation (due to the nonlinear absorption) in a waveguide in the IMOS platform based on earlier studies is of paramount importance for device design.\par

Here, we based our theoretical study on the traveling wave circuit model using the in-house modelling software \enquote{PHIsim} \cite{Phisim}, with the extension of nonlinear effects (see \nameref{s-method}). We further note that the TPA coefficient is the essential parameter for the TPA-induced nonlinear losses, which is chosen to be 1.5 $\times$ 10$^{-10}$ m/W, the value characterized for the IMOS platform \cite{Gonzalez:2009ut,Latkowski:2015uw,Gil-Molina:2018us}. On the other hand, the free carrier life-time is a key parameter for FCA-induced losses. However, there is a large discrepancy between values reported so far for InP based structures (from 1 ps - 1000 ps) \cite{Black:2017ub,Thourhout:2001vm,Bente:2021th}. Nevertheless, we found that FCA contributes negligibly ($<$ 5 $\%$) to the total nonlinear losses, even for the value of 2000 ps. We therefore chose the free carrier life-time to be a conservative value of 2000 ps (higher the lifetime, larger the loss), which is believed not to affect the validity of our result.

In our model, we investigated the energy transmission of a Gaussian optical pulse having a TE\textsubscript{0} mode profile with various pulse energies in a standard IMOS passive waveguide with a length of 1 mm, which corresponds to a reasonable dimension estimated for the length of supporting photonic distribution networks for AOS. We also considered different pulse durations in our simulation, keeping values within the margins for successful AOS \cite{Gorchon:2016aa,Wei:2021ui}. The physical parameters involved are taken from both the available literature as well as our simulations results from the MODE solver \cite{mode} (See Method). \par

            \begin{figure}[bh!]
             \centering
                \begin{subfigure}[h!]{0.5\textwidth}
                \centering
                 \includegraphics[scale = 0.3]{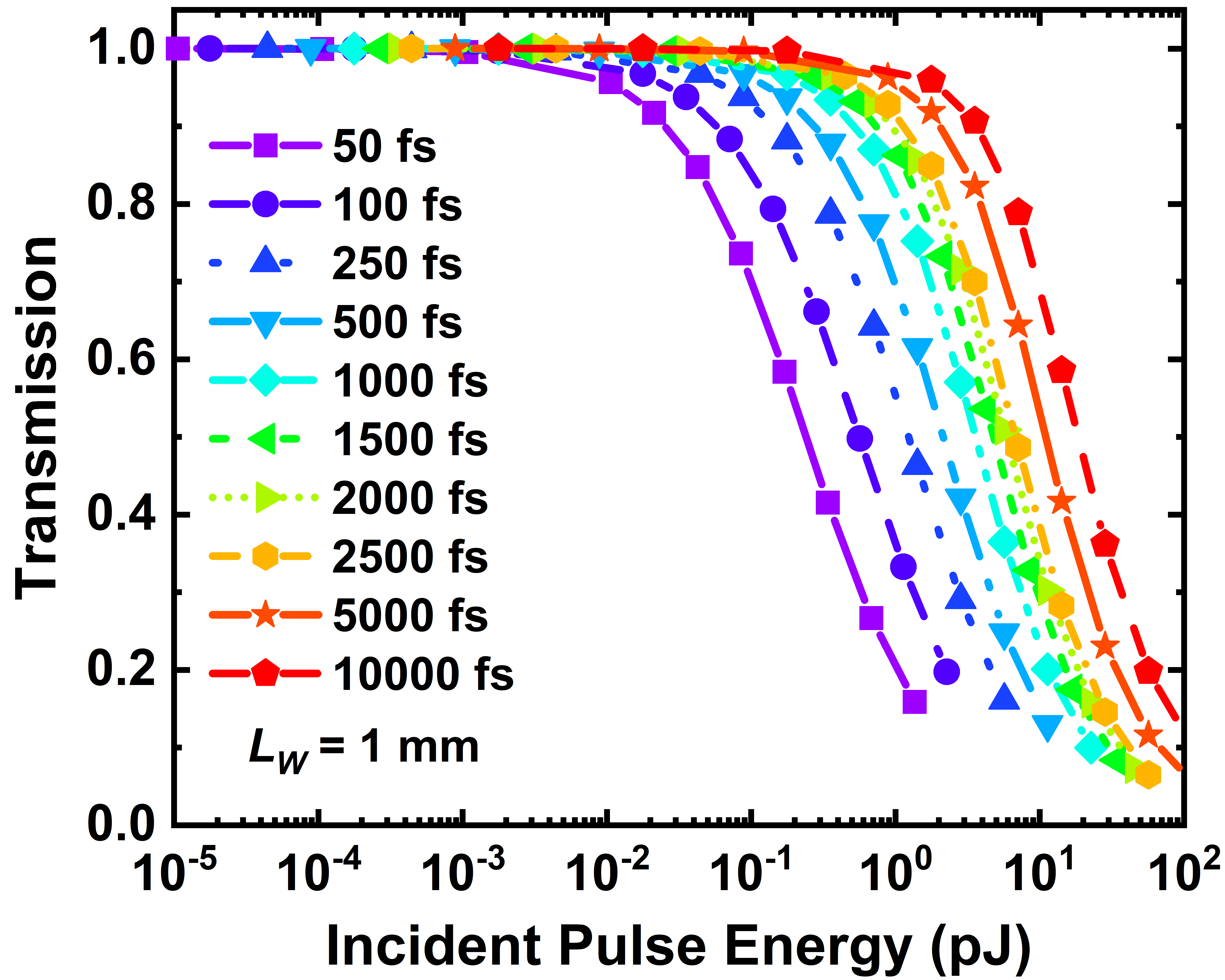}
                 \caption{}
                   \label{fig 2a}
                 \end{subfigure}
                 \begin{subfigure}[h!]{0.5\textwidth}
                \centering
                \includegraphics[scale = 0.3]{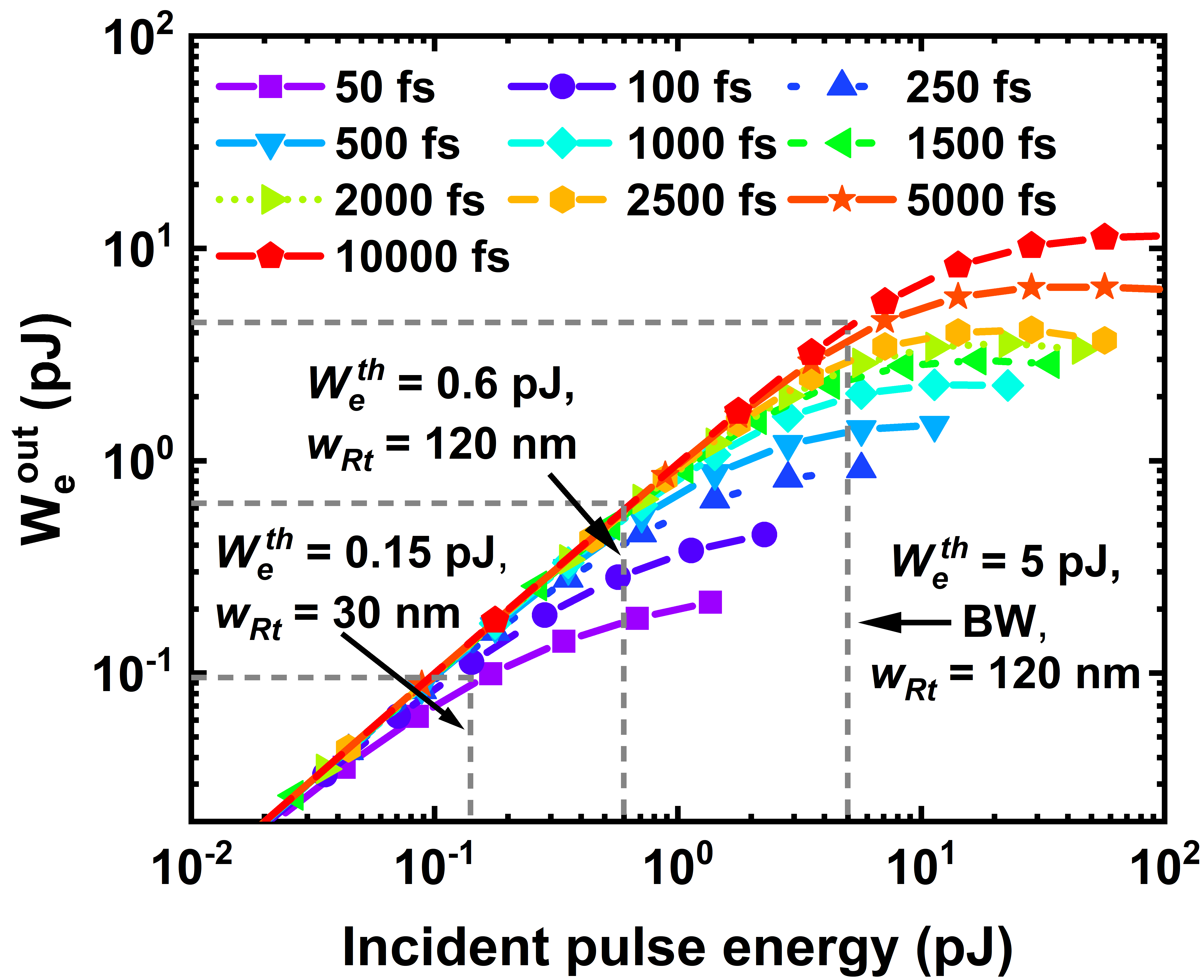}
                \caption{}
                 \label{fig 2b}
                \end{subfigure}
                 \centering
                \caption{Numerical investigation of the impact of the nonlinear absorption on AOS in the IMOS platform using the traveling wave model \enquote{PHIsim} \cite{Phisim}. (a, b) Transmission of an optical pulse and an output energy, {$W$\textsubscript{e}}\textsuperscript{out}, in terms of an incident pulse energy for various pulse durations through a 1 mm long waveguide. The gray dashed lines in (b) relate the threshold incident input pulse energies ({$W$\textsubscript{e}}\textsuperscript{th}) and their corresponding output energies, {$W$\textsubscript{e}}\textsuperscript{out}, for three different cases of the bare waveguide (BW) for a racetrack width of $w$\textsubscript{Rt} = 120 nm and the hybrid device with $w$\textsubscript{Rt} = 120 and 30 nm.}
                \label{fig 2}
                \end{figure}

As for discussing the generic insight, we first neglect the passive loss of the waveguide. Figure \ref{fig 2a} shows the transmission through the waveguide in which a \enquote{cut-off} behaviour can be seen as the energy ramps up over a threshold energy. Such a threshold energy is almost 100$\times$ higher for a 10$^4$ fs pulse than a 100 fs one, due to the fact that the \enquote{cut-off} threshold is limited by the peak power. In other words, the output energy will saturate over a certain input energy level. This suggests that in order to ensure the power efficiency for AOS and avoid possible heat dissipation due to the nonlinear absorption, the incident pulse energy needs to be kept below 15 pJ for a 10 ps pulse (see Figs. \ref{fig 2a} and \ref{fig 2b}). As for short pulses ($<$ 500 fs), such a requirement is limited to $<$ 1 pJ. The output pulse energy ({$W$\textsubscript{e}}\textsuperscript{out}) with consideration of a typical passive loss of 3.5 dB/cm for an InP waveguide in the IMOS platform \cite{van2019inp,jiao2020indium} is plotted in Fig. \ref{fig 2b}. Results show that the output energy remains linear relative to the input energy up to the \enquote{cut-off} energy point. Based on this figure, the maximum deliverable energy is $\sim$11 pJ for a pulse duration of 10 ps passing through a photonic network of 1 mm. Figure \ref{fig 2b} will serve as a benchmark in this paper, which helps to assess the performance and energy efficiency of our hybrid platform for AOS. \par

\section*{Results} \label{s-Results}

\noindent \textbf{All-optical switching.} As AOS is a heat driven process \cite{Mentink:2012aa, Ostler:2012aa}, we assess AOS in our numerical study using FDTD \cite{fdtd} based on the instantaneous absorbed energy density of the magnetic racetrack on top of the waveguide. In our simulation, we extract the spatial distribution of the absorbed energy by the racetrack, which is then compared with experimentally characterized threshold fluence of 0.5 mJ/cm\textsuperscript{2} to determine the presence of AOS. The optimal PNA-PhC device is already presented in Fig. \ref{fig 1}. Nonetheless, to show our design protocol for reaching our optimized configuration, we present the results progressively for case-studies of five configurations involving the intermediate steps, each of which was optimized, individually.\par

                
Figure \ref{fig 3a} shows the five configurations along side their corresponding two dimensional (2D) electric field distribution across the waveguide in the \textbf{XY} plane. We start by examining the case of the bare waveguide (BW) as shown in panel \RomanNumeralCaps{1}, where no other photonic component is present. In this case, the light interaction with the spintronic material is limited due to the diffraction limit, which can be evidenced by the field distribution showing almost no modification of the incident light after passing through the racetrack. As a result of the insignificant light-matter interaction, $\sim$90$\%$ of the propagating light is delivered to the waveguide's output. We further calculated the absorbed energy density in the racetrack and plotted the results along the \textbf{Y} axis by averaging the value along the \textbf{X} and \textbf{Z} axes within the region of the racetrack right on top of the waveguide as shown in Fig. \ref{fig 3b}, where the peak energy density for the case of BW is normalized as 1. The nearly flat absorption profile (see the purple curve in Fig. \ref{fig 3b}) shows the large footprint of the switching region, which almost fits the width of the waveguide. In order to exceed the energy threshold of 0.5 mJ/cm\textsuperscript{2} \cite{li2021ultra} for AOS, we found that an incident threshold pulse energy of {$W$\textsubscript{e}}\textsuperscript{th} = 5 pJ is at least required in this case. This amount of incident pulse energy is found to be unacceptable (let alone the large switching footprint) as indicated by the results shown in Fig. \ref{fig 2b} (indicated by the rightmost gray dashed line), which shows nonlinear absorption loss of ({$W$\textsubscript{e}}\textsuperscript{th} - {$W$\textsubscript{e}}\textsuperscript{out})/({$W$\textsubscript{e}}\textsuperscript{th}) $\geq$ 16$\%$ for a pulse duration of 10 ps propagating over 1 mm, caused by TPA and FCA. For a pulse duration shorter than 5 ps, such a value cannot be delivered. Considering the initial result, a certain photonic design is needed to enhance the absorbed energy density of the racetrack to minimize the nonlinear absorption loss as well as to reduce the footprint of the 
switching area. \par

For the second configuration, we coupled the racetrack with a PNA as shown in panel \RomanNumeralCaps{2} of Fig. \ref{fig 3a}. Based on the electric field distribution in this panel, a slight enhancement in the light-matter interaction can be seen, which resulted in a waveguide transmission of 75$\%$. However, this enhancement is not sufficient to provide enough energy for AOS such that according to Fig. \ref{fig 3b}, the maximum absorbed energy by the racetrack is enhanced only from 1$\%$ to 1.9$\%$. To further enhance the absorption cross section of the racetrack, we created a configuration based on a one dimensional periodic array of PNAs (panel \RomanNumeralCaps{3} in Fig. \ref{fig 3a}). By making an array of PNAs with a certain pitch, plasmonic surface lattice resonance (PSLR) is excited via coupling LSPR of each PNA together \cite{kataja2015surface,kravets2018plasmonic}. The resultant PSLR can further increase the localized electric field and enhance the absorbed energy density of a nanoparticle coupled to the PNA array. The light transmission in this case drops from 75$\%$ to $\sim$48$\%$, implying further enhancement in the interaction between the propagating light, PNA array, and racetrack. By looking at Fig. \ref{fig 3b}, we see a pronounced spike in the absorbed energy density having a maximum enhancement of 3.2$\times$ with a full width at half maximum (FWHM) of $\sim$140 nm. Despite the drop in the transmission for the case of the PNA array, we see that the absorbed energy density of the racetrack did not increase significantly. Addition of more PNA elements in the array did not improve the result due to the lossy nature of gold \cite{johnson1972optical}.\par

            \begin{figure*}[hb!]
             \centering
                \begin{subfigure}[ht!]{\textwidth}
                \centering
               \includegraphics[scale = 0.6]{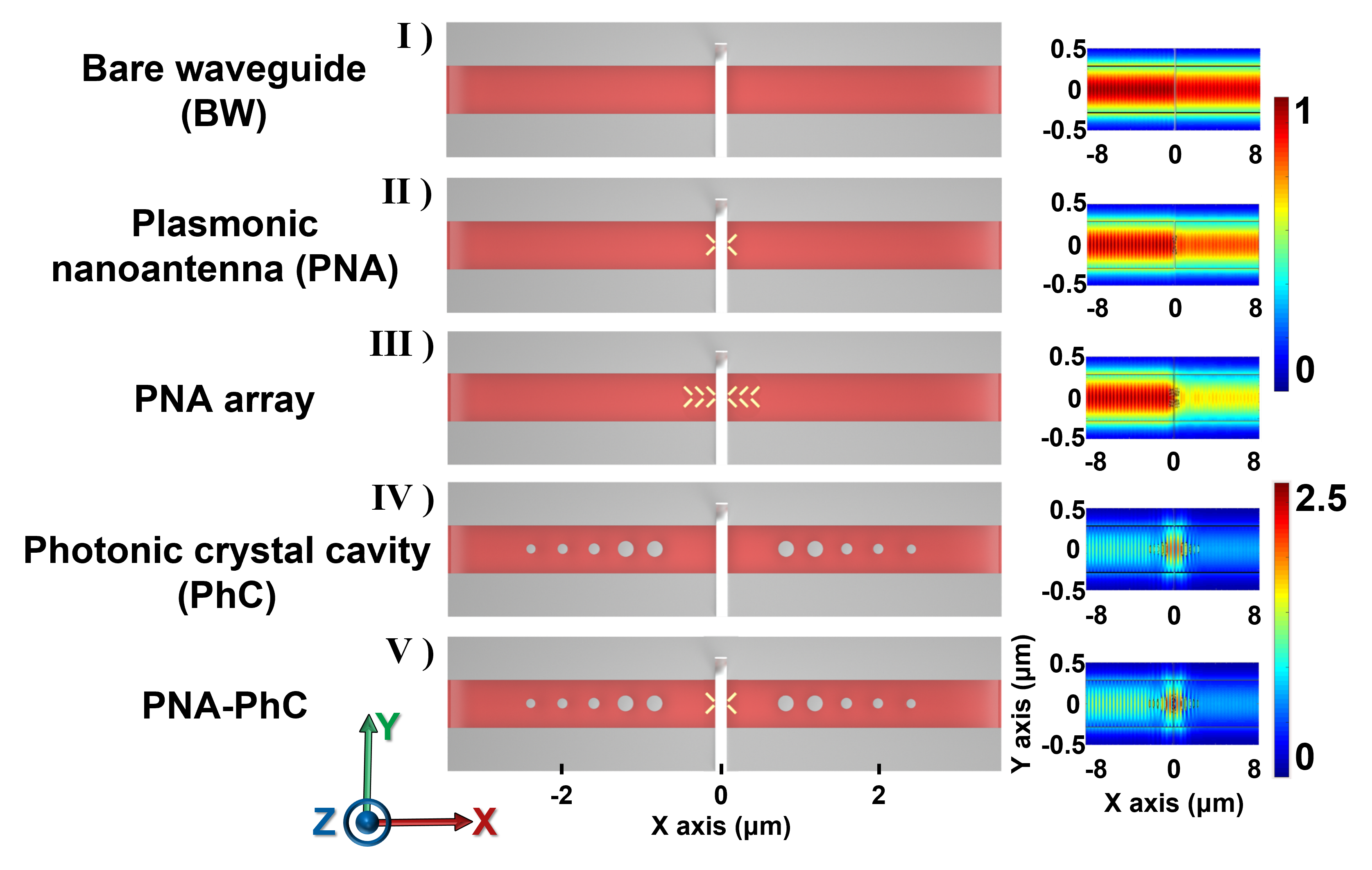}
                 \caption{}
                   \label{fig 3a}
                 \end{subfigure}
                 \begin{subfigure}[t]{0.49\textwidth}
                \centering
                \includegraphics[scale = 0.3]{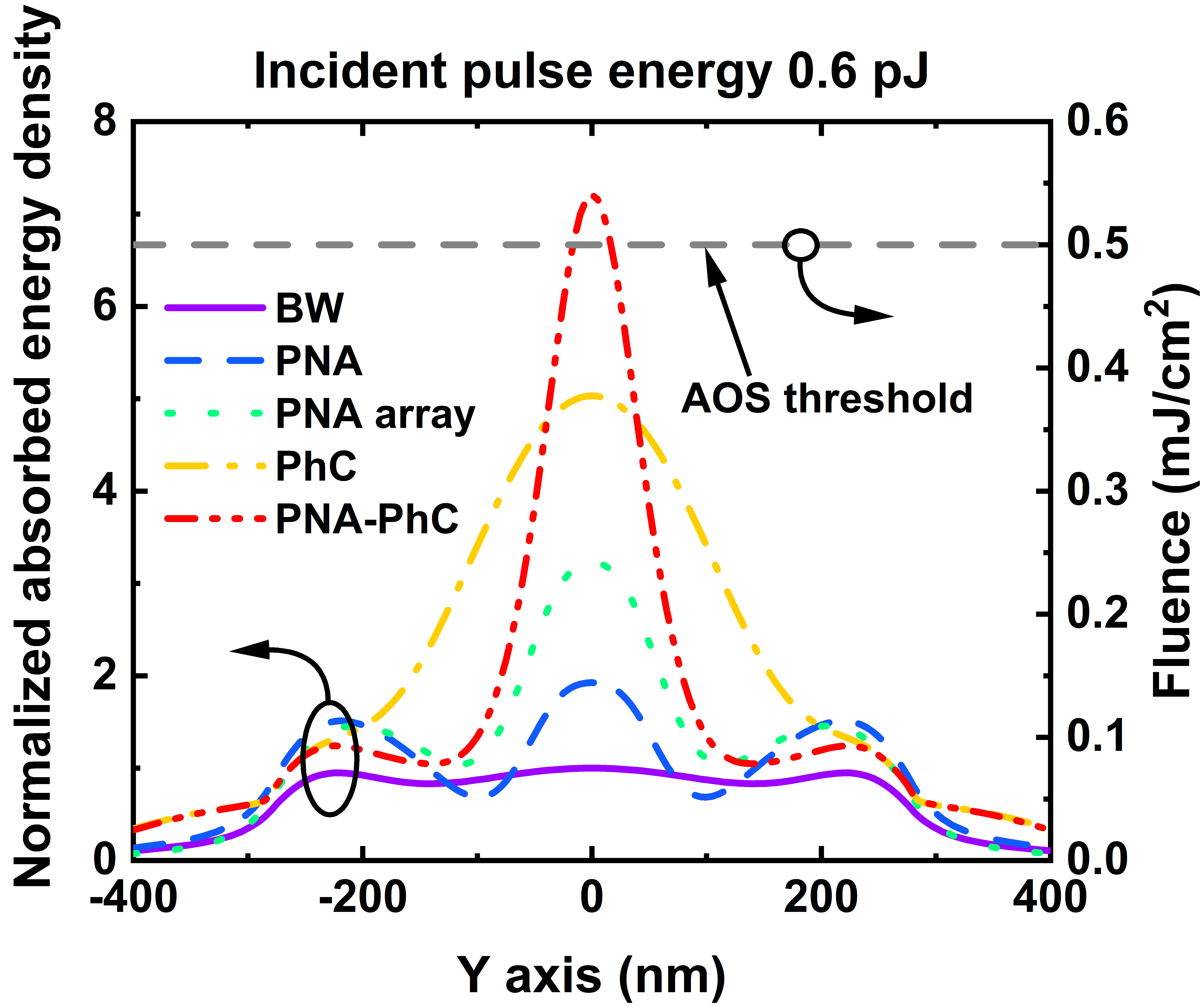}
                \caption{}
                 \label{fig 3b}
                \end{subfigure}
                \begin{subfigure}[t]{0.49\textwidth}
                \centering
                 \includegraphics[scale = 0.3]{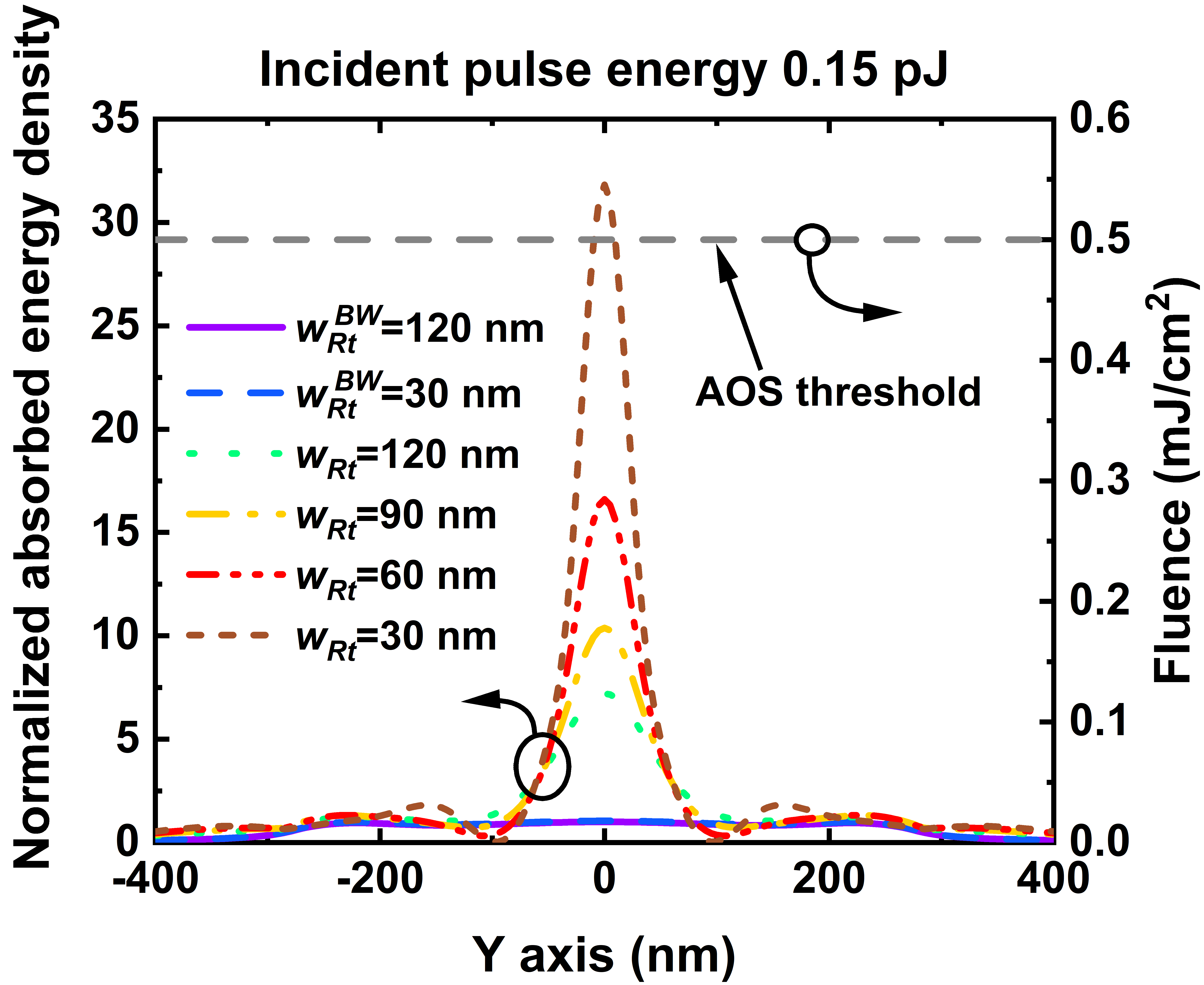}
                  \caption{}
                  \label{fig 3c}
                 \end{subfigure}
                  \centering
                \caption{Illustration of AOS. (a) Schematic diagram of five different configurations alongside their correspondent electric field distributions in the \textbf{XY} plane through the waveguide. Normalized absorbed energy density and the projected fluence, averaged along the width and thickness of the racetrack for: (b) five different configurations, and (c) the PNA-PhC device (\RomanNumeralCaps{5}) with racetrack widths ($w$\textsubscript{Rt}) of 30 to 120 nm as well as the BW device (\RomanNumeralCaps{1}) with $w$\textsubscript{Rt} of 120 and 30 nm, respectively.}
                \label{fig 3}
                \end{figure*}

Based on the results so far, we looked for alternative approaches to find a solution that brings further enhancement in the energy absorption by the racetrack. As mentioned in section \nameref{s-Intro}, a PhC cavity is another photonic component that can enhance the electric field by spatial confinement of light. So, we designed another configuration based on a PhC cavity as schematically presented at the left side of panel \RomanNumeralCaps{4} in Fig. \ref{fig 3a}. The field distribution plot at the right side of this panel clearly shows the appearance of the cavity mode from which the electric field enhancement of 2.5$\times$ in the middle of the waveguide is achieved. For this configuration, the light transmission is $\sim$54$\%$. For the design of the cavity careful attention was paid to the reflectivity which is only $\sim$8$\%$ and makes such a design highly energy efficient. Lowering the reflection was done firstly by reducing the number of holes as well as tapering down the radii of the holes toward the end sides of this cavity to reduce the mismatch between the waveguide and cavity modes as much as possible, while keeping the electric field enhancement reasonably high (see panel \RomanNumeralCaps{4} in Fig. \ref{fig 3a}). According to Fig. \ref{fig 3b}, introducing the configuration with the PhC cavity is promising in that the absorbed energy density shows an enhancement of 1.6$\times$ relative to the PNA array configuration, i.e. the peak absorbed energy density rises from 3.2 to 5, while in general it shows 5$\times$ enhancement relative to the BW configuration (see Fig. \ref{fig 3b}). However, the drawback of this configuration is its broad absorption profile (with a FWHM of $\sim$260 nm) across the racetrack, which is originated from the physical issue that a PhC cavity cannot overcome the diffraction limit of light. Upon increasing the incident pulse energy, the broad absorption profile of this configuration can increase the probability of the occurrence of TPA and FCA at the center of the cavity, which consequently can prohibit AOS to happen.\par 

In order to take advantage of the higher power absorption enhancement of the PhC configuration and the narrower FWHM of the PNA one, we propose the hybrid scheme in which a PNA is coupled at the center of the PhC cavity as shown in panel \RomanNumeralCaps{5} of Fig. \ref{fig 3a}. The reason for using one PNA instead of an array is to avoid additional light loss in the system by the extra PNA elements constituting the array. Using this configuration, the light transmission decreases further from 54$\%$ to 43$\%$, while the reflection is as small as $\sim$12$\%$. Based on Fig. \ref{fig 3b}, the hybrid PNA-PhC configuration resulted in a maximum 7$\times$ enhancement in the absorbed energy density compared to the initial BW configuration. More importantly, using this configuration, the maximum amount of power absorbed by the racetrack is high enough such that for an incident pulse energy of 0.6 pJ, we reached the threshold fluence of 0.5 mJ/cm\textsuperscript{2} for AOS. 
Referring back to Fig \ref{fig 2b} shows that for an input power of 0.6 fJ, the relation between the output and input energies is in the linear regime, which implies that the probability of the occurrence of TPA and FCA is suppressed significantly. \par

With the advancement in spintronic applications, an MRAM size down to sub 100 nm with a robust operation has long been demonstrated \cite{Dieny:2020ub}. Moreover, the interfacial anisotropy ($>$ 0.5 MJ/m$^2$) in Pt/Co/Gd \cite{Luding:2022ur, Wang:2020ab,Li2022He} is expected to be enough to support thermally stable sub 100 nm domains. It is a known fact that scaling down MRAM and racetrack brings further energy/speed gain \cite{Garello:2019aa, blasing2020magnetic}. Hence, we inspected the impact of down-scaling of the racetrack's width, $w$\textsubscript{Rt}, on the AOS performance with the PNA-PhC device. Here, we keep a fixed distance between the edge of the racetrack and PNA. The absorbed energy density is plotted for $w$\textsubscript{Rt} of 120, 90, 60, and 30 nm in Fig. \ref{fig 3c}. For comparison purpose, the results for the BW configuration with $w$\textsubscript{Rt} of 120 and 30 nm are also plotted. The hybrid device with $w$\textsubscript{Rt} = 30 nm achieved an enhancement of $\sim$32$\%$ compared to the BW one with $w$\textsubscript{Rt} of 120 and 30 nm. This enhancement is related to the stronger light-matter interaction due to the reduced distance between the PNA elements. More interestingly, for such a large enhancement of the absorbed energy density by the hybrid device at $w$\textsubscript{Rt} = 30 nm, the threshold fluence can be reached for an incident pulse energy as small as 0.15 pJ. Furthermore, by looking at Fig. \ref{fig 2b}, one can see that for pulse lengths of 50 to 10$^4$ fs, the system's response for a 0.15 pJ input pulse energy is linear, pointing to the absence of the TPA and FCA phenomena in the device. Moreover, the absorbed energy density was found to have a FWHM of $\sim$60 nm, much smaller than the conventional optics are able to focus. Therefore, our results elucidate the capability of the hybrid design in optically addressing spintronic devices of sub 100 nm, compatible for high density data packing. \par

\vline


\noindent \textbf{All-optical reading.} So far, we showed the possibility of AOS on-chip in an area/energy efficient fashion with the help of the combined efforts from both PNA and PhC. Similarly, MO interactions can be likewise enhanced using the hybrid device. Following switching a magnetic state, the magnetic information can be read out on the same photonic chip using PMOKE by which information can be encoded in the photonic domain as an intensity variation of the transverse magnetic (TM\textsubscript{0}) mode. We will evaluate the performance of AOR based on the magnitude of the PMOKE induced polarization change (Kerr rotation) and the minimum bit size that can be distinguished independent of all other neighboring bits in the racetrack. In order to find the minimum bit size readable in the racetrack, we investigate the evolution of the Kerr rotation along the propagation direction of light through the waveguide in terms of the size of the target magnetic bit in the presence of the oppositely magnetized rest of the racetrack. Then, from this information, we calculate the magnitude and phase of the Kerr rotation in terms of the target bit size to explore the minimum readable bit size.\par

Figure \ref{fig 4a} shows the evolution of the Kerr rotation across the waveguide of PNA-PhC, Fig. \ref{fig 3a}, in terms of different domain widths (DWs) of the target magnetic bit. The target magnetic bit is shown in red color surrounded by the oppositely magnetized background in green as sketched in the inset. As light interacts with the magnetic racetrack, a rise in the polarization rotation, i.e. Kerr rotation, as a result of PMOKE can be seen. The waveguide medium induces birefringence between the TE\textsubscript{0} and the PMOKE-induced TM\textsubscript{0} components of light as it propagates through the waveguide, leading to a beating between the two components. As a result, the induced Kerr rotation will be superposed with the beating oscillation, which in turn leads to an oscillation in the Kerr rotation along the light propagation direction as shown Fig. \ref{fig 4a}. Note that the small oscillation in the Kerr rotation before the magnetic racetrack is originated from the light back-reflected from the racetrack and PNA. According to Fig. \ref{fig 4a}, the oscillation of the Kerr rotation, $\theta$\textsubscript{K}, for DW = 60 nm, is out of phase relative to the case with larger DWs such that it is almost reversed by 180$^\circ$ relative to the larger DWs.\par

To elaborate on this phase reversal, it is worth pointing out that the Kerr rotation is the sum of the contributions from all magnetic regions, i.e. the target bit and the background regions. For small DWs, the MO response originates mostly from the oppositely magnetized background regions based on which we cannot detect the magnetization state of the target bit. On the other hand, as DW increases from 120 to 570 nm, the MO contribution from the target bit dominates the MO contributions from the oppositely magnetized background regions, which consequently makes the magnitude of the Kerr rotation of the target bit greater than the superposition of the Kerr rotations of the background regions. Therefore, the magnetization state of the target bit can be identified unambiguously. A slight phase shift between the cases is a consequence of the phase difference in the beating pattern between the TE\textsubscript{0} and PMOKE-induced TM\textsubscript{0} components. Thus, we used the magnitude and phase of the Kerr rotation to calculate the strength of the MO interaction and the minimum readable bit size. \par

             \begin{figure}[b!]
             \centering
              \begin{subfigure}[h!]{0.49\textwidth}
               \includegraphics[scale=0.29]{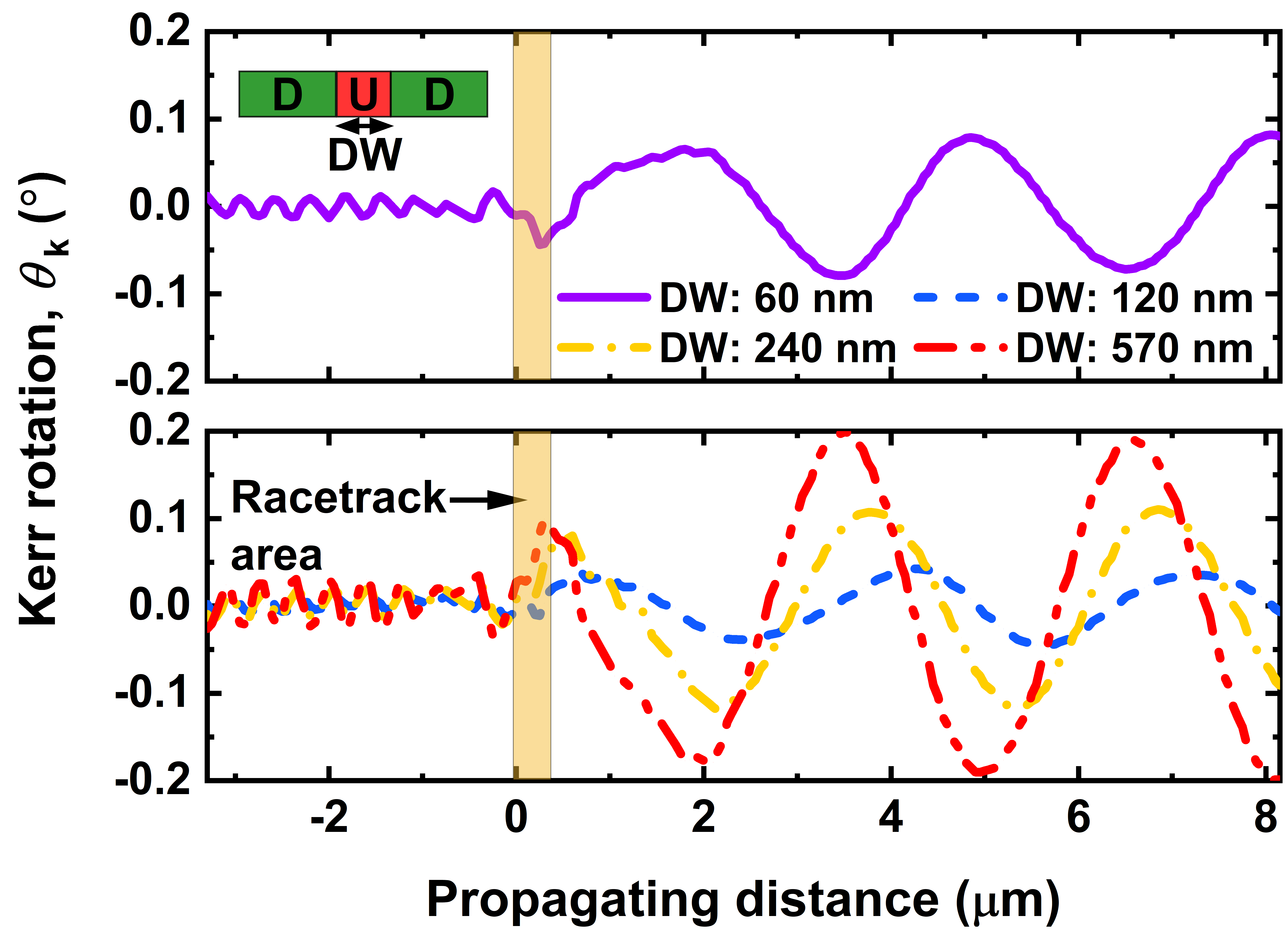}
                  \caption{}
                    \label{fig 4a}
                    \end{subfigure}
                    \centering
                    \caption{Illustration of AOR. (a) The evolution of the Kerr rotation, $\theta$\textsubscript{K}, along the propagation direction inside the waveguide of the PNA-PhC device in terms of target magnetic bit widths of DW = 60, 120, 240, and 570 nm. The inset shows the target bit in red surrounded by oppositely magnetized background in green.}
                    \end{figure}

            \begin{figure}[b!]
            \ContinuedFloat
              \begin{subfigure}[h!]{0.49\textwidth}
                \centering
                 \includegraphics[scale=0.29]{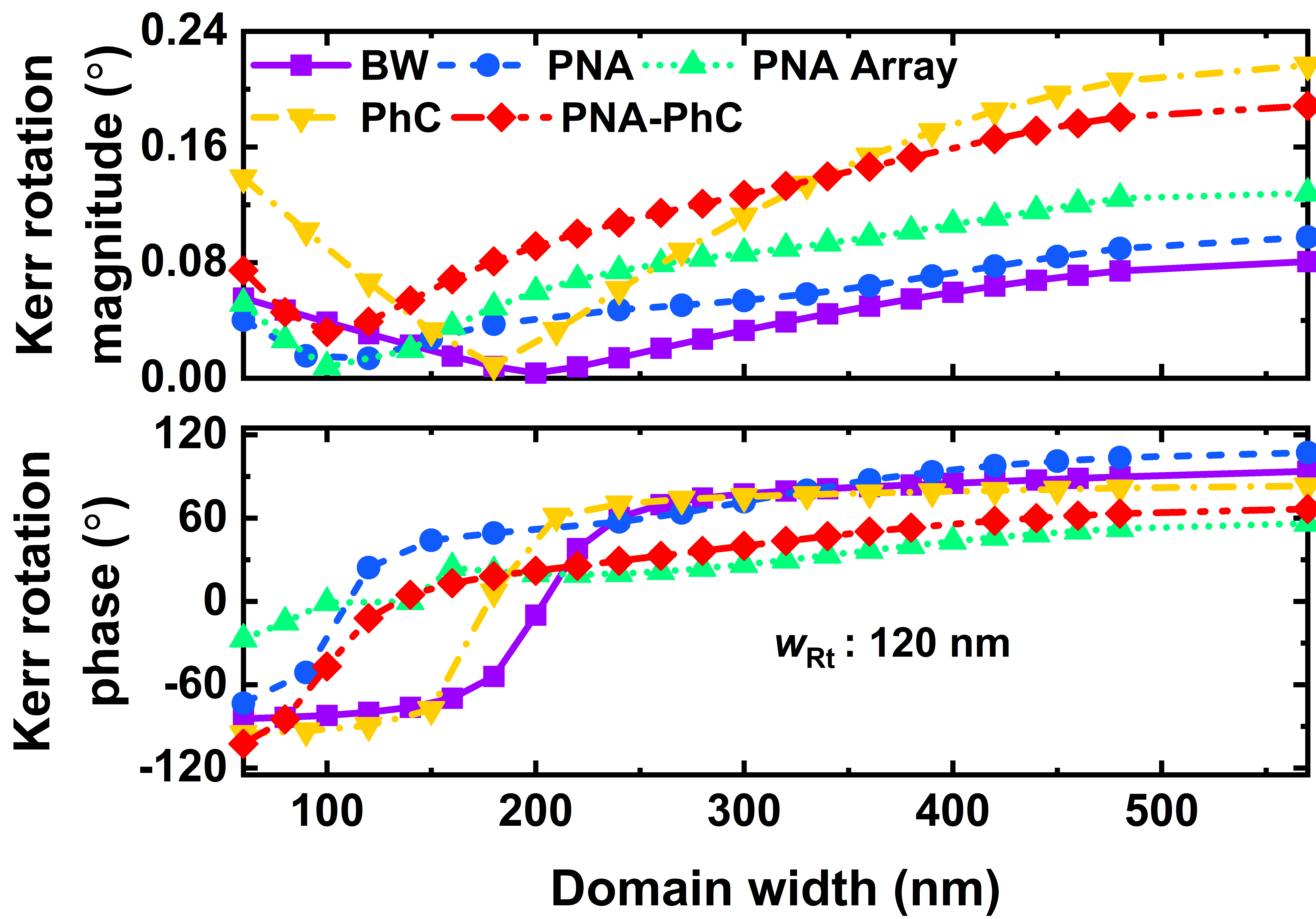}
                  \caption{}
                  \label{fig 4b}
                 \end{subfigure}
                 \begin{subfigure}[h!]{0.49\textwidth}
                    \centering
                  \includegraphics[scale=0.29]{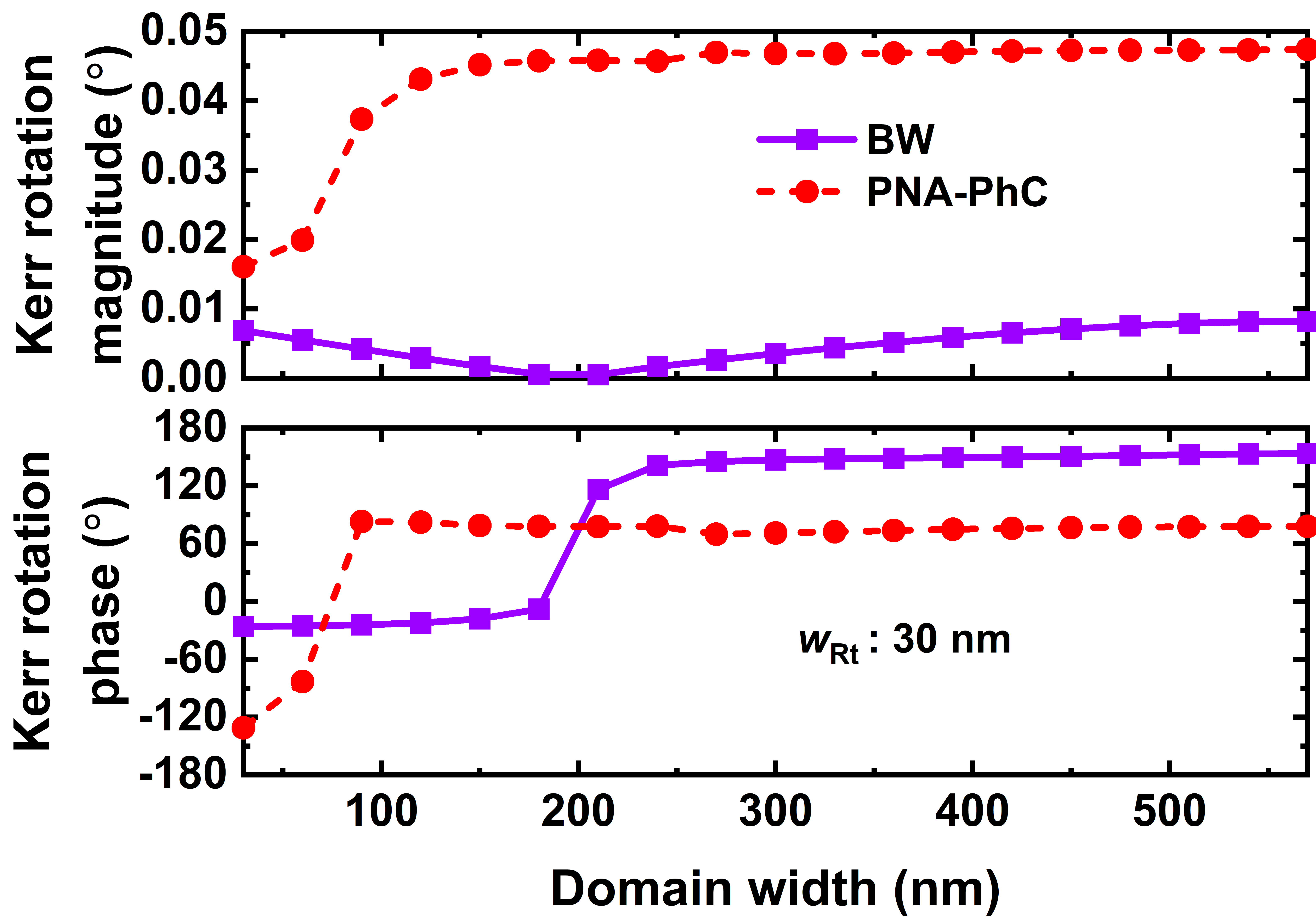}
                    \caption{}
                     \label{fig 4c}
                    \end{subfigure}
                    \centering
                    \caption{Continued. Illustration of AOR. The Kerr rotation magnitude and phase in terms of the target DW for: (b) the five different configurations with \textit{w}\textsubscript{Rt} of 120 nm, and (c) the BW and PNA-PhC devices with \textit{w}\textsubscript{Rt} of 30 nm. The inset shows the target bit in red surrounded by oppositely magnetized background in green.}
                    \label{fig 4}
                    \end{figure}

We extracted both the magnitude and phase of the Kerr rotation for all five configurations with \textit{w}\textsubscript{Rt} of 120 nm (see Figure \ref{fig 3a}) as a function of DW, where the results are plotted in Fig. \ref{fig 4b}. By increasing DW, there is a minimum in the magnitude of the Kerr rotation which is accompanied by a sudden transition in the phase of the Kerr rotation in all devices. The minimum Kerr rotation happens at DW = 100 nm (200 nm) for the PNA-PhC (BW) device which comes with a jump in the state of the Kerr rotation phase as shown in Fig. \ref{fig 4b}. The minimum Kerr rotation magnitude and the jump in the phase are indeed due to the destructive interference between the individual Kerr rotations of the target bit and the oppositely magnetized neighboring bits in the racetrack. For very small DWs, e.g. DW $<$ 100 nm (200 nm) in the PNA-PhC (BW) device, the PMOKE response of the target bit is much smaller than the superposition of the PMOKE responses from the background regions with the opposite magnetization due to the limited MO contribution. Therefore, the resultant PMOKE response is determined dominantly by the magnetization in the rest of the racetrack. In contrast, with increasing DW, e.g. DW $>$ 100 nm (200 nm) in the PNA-PhC (BW) device, due to the continual enhancement of the MO contribution from the target bit, the PMOKE response from this bit gradually increases and dominates the PMOKE response from the superposition of the oppositely magnetized background bits. So, the target magnetic bit with the ‘up’ magnetization can be unambiguously determined above this value of DW, independent of the bit pattern in the background regions.\par

The threshold DW for which the magnetization in the target bit can be detected is considered a measure for determining the resolution of the device. Based on Fig. \ref{fig 4b}, we can see that the PhC device improved the PMOKE response relative to the BW device, but it did not enhance the resolution due to the diffraction limit. However, the devices based on the PNA which benefit from the LSPR increased the resolution by $\sim$100 nm through enhancing the effective polarizability of the target bit in the subwavelength regime.\par

Finally, we compared the AOR performance of the BW and hybrid configurations for \textit{w}\textsubscript{Rt} of 120 nm and 30 nm to see the impact of shrinking the width of the racetrack on the AOR function. Based on Fig. \ref{fig 4c}, we can see that the detection resolution of the BW device does not change. In contrast, for the case of the hybrid device, due to the enhanced interaction between the PNA elements, the detection resolution reaches to below 100 nm. Furthermore, by comparing the strength of the Kerr rotation in both cases, it is clear that the hybrid device has much better performance compared to the BW one with $\sim$5$\times$ enhancement for DWs $>$ 100 nm. Overall, this section showed the possibility of determining the magnetization state in magnetic bits sizes of $\sim$ 100 nm for the racetrack widths between $w$\textsubscript{Rt} = 30 and 120 nm, regardless of the magnetization state in the rest of the racetrack, using the hybrid PNA-PhC device. \par

\vline


\noindent \textbf{Detection of the magnetization state.} Reversing the magnetization state inverts the rotation angle of the polarization of the TE\textsubscript{0} component due to PMOKE. Consequently, this reversal in the rotation angle makes the phase of the PMOKE-induced TM\textsubscript{0} component relative to the TE\textsubscript{0} component changes by 180$^\circ$. To be able to detect this phase variation, a necessary next step is to convert such a response into a reading signal. In this section, we adopt the method proposed by Demirer \textit{et al.} to convert this phase change to an intensity variation of the TM\textsubscript{0} component \cite{demirer2021magneto,demirer2022integrated}. \par
 
Here, we introduce a polarization converter (PC) based on an integrated plasmonic quarter-wave plate (QWP) \cite{gao2015chip} as an essential building block for such a conversion. Compared to conventional PCs based on slanted waveguides \cite{deng2005design,demirer2021magneto,demirer2022integrated}, a plasmonic PC has a higher fabrication process tolerance and we can tune the device performance by easily adjusting its geometrical parameters \cite{gao2013ultra}.\par

As the wave propagates through the PC, the phase difference can be translated to an intensity variation because of the partial conversion of the TE\textsubscript{0} component to the TM\textsubscript{0} component, where the intensity variation can be detected using an on-chip photodetector. The schematic of the PC is depicted in Fig. \ref{fig 5a} in which for the sake of clarity, the surrounding SiO\textsubscript{2} material is omitted. A gold metal layer with a thickness of 30 nm is coupled onto the InP waveguide. The length of the PC is $l$\textsubscript{PC} = 1860 nm, while its width ($w$\textsubscript{PC}) is 180 nm, respectively. To avoid excessive absorption loss caused by the gold, a SiO\textsubscript{2} spacer layer with a thickness of 20 nm, so $h$\textsubscript{PC} = 50 nm, and with the same length is introduced between the InP waveguide and the gold thin film. It is important to note that the dimensions of the proposed PC are optimized to create efficient QWP functionality, i.e. rotating the eigenstates by 45$^\circ$ and making the difference in the magnitude of the TE\textsubscript{0} and TM\textsubscript{0} components approach zero.\par

To elaborate on the process of translating the phase change to the intensity variation using the PC for up and down magnetization states, we plotted the polarization states of the propagating light before, through and after the PC using the Poincar\'{e} sphere for the hybrid device and the racetrack width of $w$\textsubscript{Rt} = 120 nm, where the target magnetic bit has a DW = 200 nm. This specific value is chosen according to Fig. \ref{fig 4b} in which the Kerr rotation magnitude is almost zero for the BW device. For this purpose, the Stokes parameters $S_1$ to $S_3$ are defined as follows:

                \begin{equation}
                    S_1 = \cos{2\chi} \cos{2\psi},
                    \label{Eq-1}
                \end{equation}
                \begin{equation}
                    S_2 = \cos{2\chi} \sin{2\psi},
                \end{equation}
                \begin{equation}
                    S_3 = \sin{2\chi},
                \end{equation}
                
\noindent where $\psi$ and $\chi$ are the polarization rotation angle (Kerr rotation) and ellipticity angle (Kerr ellipticity), $S_1$ relates to the linearly polarized TM\textsubscript{0} and TE\textsubscript{0} components of light, $S_2$ indicates the orientation of linearly polarized modes at $\pm45^\circ$, and $S_3$ shows the light is whether left- or right-hand side circularly polarized light \cite{collett2005field}.\par

            \begin{figure}[b!]
             \centering
                \begin{subfigure}[h!]{0.49\textwidth}
                \centering
                \includegraphics[scale=0.45]{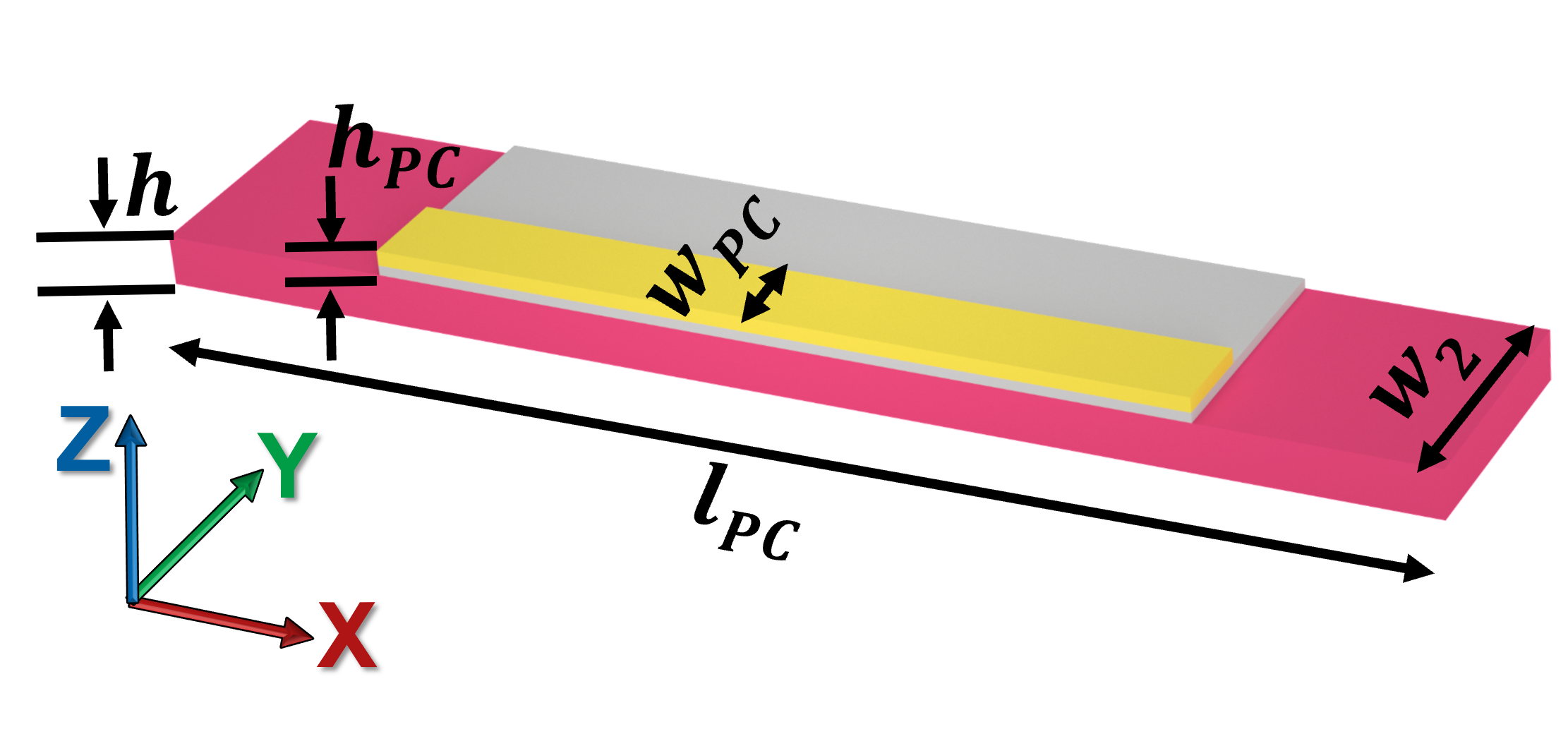}
                 \caption{}
                   \label{fig 5a}
                   \end{subfigure}
                   \centering
                    \caption{Translation of a phase change to an intensity variation. (a) Schematic of the hybrid polarization converter (PC), where $w$\textsubscript{2}, $w$\textsubscript{PC}, $h$, $h$\textsubscript{PC}, and $l$\textsubscript{PC} are widths of the InP waveguide and PC, the height of the waveguide and PC, and the length of PC.}
                    \end{figure}

    \begin{figure}[hb!]
        \ContinuedFloat
             \centering
               \begin{subfigure}[h!]{0.49\textwidth}
                \centering
                 \includegraphics[scale=0.42]{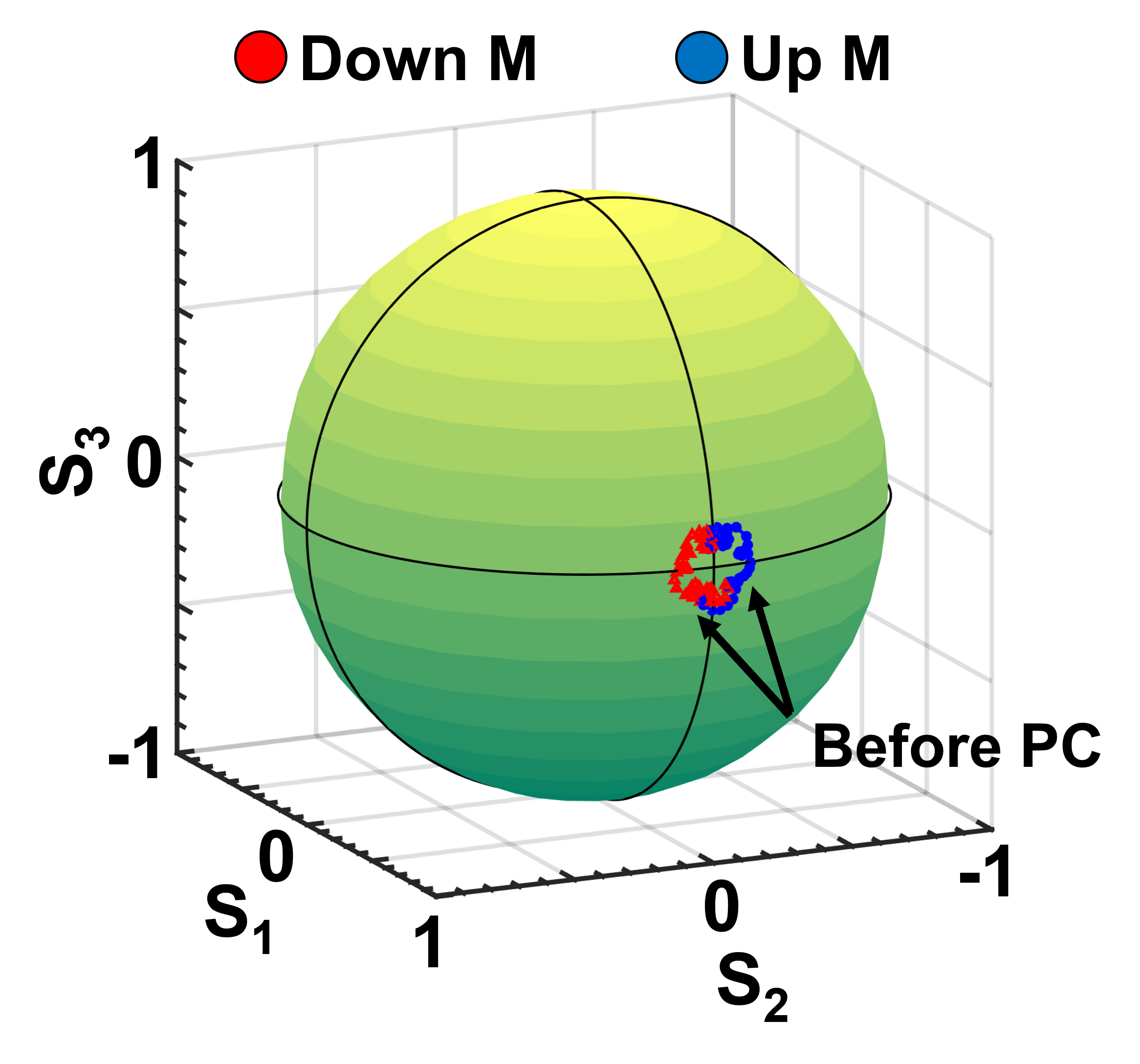}
                  \caption{}
                  \label{fig 5c}
                 \end{subfigure}
                 \begin{subfigure}[h!]{0.49\textwidth}
                \centering
                \includegraphics[scale=0.42]{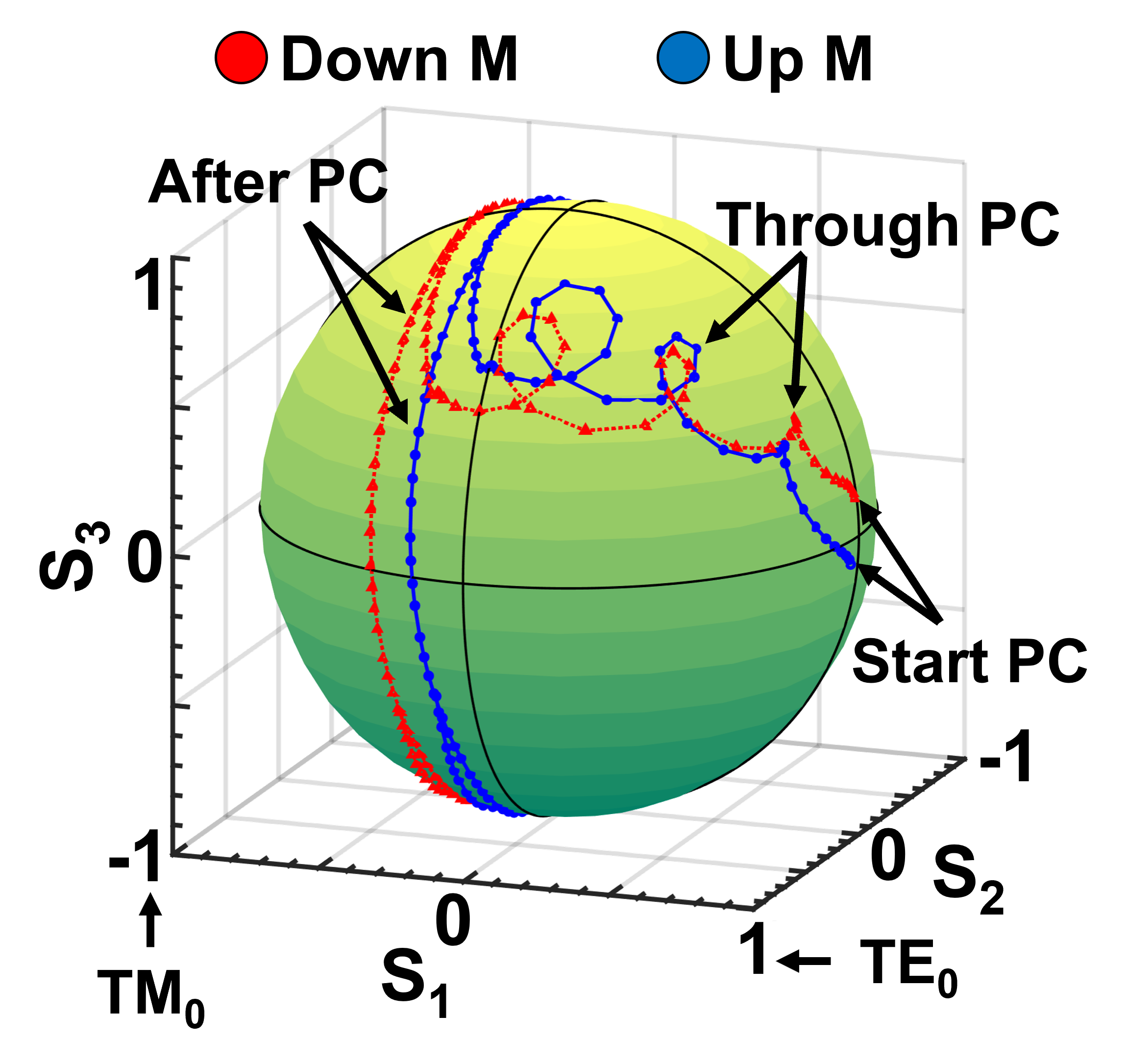}
                  \caption{}
                  \label{fig 5d}
                 \end{subfigure}
                   \centering
                    \caption{Continued. Translation of a phase change to an intensity variation. Polarization state of light propagating through the waveguide using the Poincar\'{e} sphere: (b) before, (c) through and after PC for a racetrack width of $w$\textsubscript{Rt} = 120 nm and the target magnetic bit with DW = 200 nm. Results for up (red) and down (blue) magnetization states are shown for a exaggerated value of MO efficiency for the clarification.}
                    \label{fig 5}
                    \end{figure}

According to Fig. \ref{fig 5c}, while the propagating light has not entered the PC yet, the Stokes parameters on the Poincar\'{e} sphere indicate that $S_1 \approx$ 1, and $S_{2, 3} \approx$ 0 upon changing the magnetization direction. According to these values, we can deduce that the propagating light has a dominant TE\textsubscript{0} component inside the waveguide before the PC, where changing the magnetization direction, induces a phase difference between the TE\textsubscript{0} and PMOKE-induced TM\textsubscript{0} components of light. However, the PMOKE effect is so small that it cannot make a notable evolution in the polarization state of the incoming light. In contrast, when the light passes through the PC, the partial polarization conversion gives rise to the emergence of the TM\textsubscript{0} component with significant magnitude. Then, both the TE\textsubscript{0} and TM\textsubscript{0} components beat together along the waveguide due to the waveguide's birefringence. This polarization conversion can be clearly seen in Fig. \ref{fig 5d}, where the polarization state of the propagating light evolves in the $S_2 \, S_3$ plane, i.e. -1 $\leq$ $S_{2,3}$ $\leq$ 1. In sharp contrast to the previous case, we observe two slightly different values of $S_1$ for the up and down magnetization directions, respectively (see the blue and red points in Fig. \ref{fig 5d}). This difference in the value of $S_1$ reflects the intensity variation in the TM\textsubscript{0} component of light originated from the PMOKE. In other words, the PMOKE-induced phase difference between the TE\textsubscript{0} and TM\textsubscript{0} components of the propagating light is translated to a PMOKE-induced variation in the intensity of the TM\textsubscript{0} component. Note that the spiral shape of the polarization evolution before and throughout the PC is because of the reflection and scattering off the gold metal layer back to the waveguide.\par

To provide memory read-out based on the intensity variation of the generated TM\textsubscript{0} component,
a figure of merit $\Delta S_1$, representing the relative contrast, is defined as follows:

                \begin{equation}
                    \Delta S_1 (\%) = \lvert S_{1, \uparrow} - S_{1, \downarrow} \rvert \times 100,
                    \label{Eq-4}
                \end{equation}
                
\noindent where $S_{1, \uparrow}$ and $S_{1, \downarrow}$ are the output $S_1$ Stoke parameters defined by Eq. (\ref{Eq-1}) for the upward and downward magnetization directions, respectively. Based on Eqs. (\ref{Eq-1}, \ref{Eq-4}), the relative contrast for the BW and the PNA-PhC devices are 0.03$\%$ and 0.64$\%$, respectively.\par

We can see that using the bare waveguide, it is not possible to detect the difference in the state of a magnetic bit with a size beyond the diffraction limit. Conversely, our hybrid device curbed the diffraction limit challenge and identified the change in the magnetization direction in the same bit in the presence of oppositely magnetized background regions. It is worth mentioning that as plotted in Figs. \ref{fig 4b} and \ref{fig 4c}, DW = 200 nm is not the smallest magnetic bit that can be read using the hybrid device. In fact, our proposed device can read magnetic bits with a footprint down to (DW $\times$ \textit{w}\textsubscript{Rt}) $\sim$100 $\times$ 100 nm\textsuperscript{2} no matter what the magnetization state of the rest of the racetrack is, pointing towards the potential of our proposed hybrid device in enhancing MO effects at the nanoscale.\par

\subsection*{Discussion} \label{s-Dis}

The hybrid integration scheme of spintronic and photonics has been a hot discussion. As for approaches for light projection onto spintronic devices, Becker \textit{et al.} \cite{Becker:2020aa} proposed a 2-D grating coupler to project light pulses vertically into AOS-switchable magneto-tunnel junctions (MTJs) \cite{Chen:2017aa,Luding:2022ur,Aviles-Felix:2020aa}. Nevertheless, the large footprint and inefficient coupling curbed the further progress of such an approach \cite{Sobolewska:2020aa}. In addition, an extra requirement for inter-wafer integration imposes further challenges. In terms of multiplexing, Kimel and Li \cite{Kimel_AOS_Review2019} proposed an integration scheme between AOS-switchable MTJs and photonic networks using cross-coupled photonic waveguides, in which the multiplexing is realized by the wavelength division multiplexing based on arrays of Bragg gratings, which dictates the direction of wavelength-dependent out-coupling. However, such a multiplexing scheme of the light control comes with drawbacks of large footprint (diffraction limited to multiple of wavelength) and low coupling efficiency. In our proposed monolithic device, the energy efficient multiplexing is enabled by ultrafast domain wall motion \cite{Li:2021wr} in a magnetic racetrack \cite{Parkin:2008aa,blasing2020magnetic}. Moreover, the racetrack itself allows for ultra-dense data packing \cite{blasing2020magnetic} (F\textsuperscript{2} $<$ 2). Our compact hybrid device further allows for photonic control of sub-100 nm spintronic devices, making our design an enabling technology for bridging integrated photonics with spintronics.\par

Our design provides further a step-beyond analysis on AOR, compared to the work presented by Demirer \textit{et al.} \cite{demirer2021magneto,demirer2022integrated}. Demirer \textit{et al.} implemented the photonic reading of a diffraction-limited magnetic cladding. However, in our design, we have considered a complex case with a domain size beyond the diffraction-limit, combined with oppositely magnetized background domains in the rest of the racetrack. Furthermore, our plasmonic PC offers the ease of fabrication as already mentioned.\par 

Our photonic reading scheme offers an alternative way of addressing magnetic information. Despite of the fact that the spintronic way of efficient reading exists (based on AOS-MTJs) \cite{Luding:2022ur,Chen:2017aa,Aviles-Felix:2020aa}, here we argue that our approach allows to encode the magnetic information in the photonic domain giving further room for photonic manipulation.\par  

\subsection*{Conclusion} \label{s-Con}

In conclusion, we designed an integrated hybrid plasmonic-photonic device capable of AOS and AOR of nanoscale ferrimagnet bits possessing PMA in a magnetic racetrack coupled onto an IMOS platform. The hybrid device is consisted of a double V-shaped PNA coupled with a PhC cavity. Enhanced, localized light offered by the strong light-mater interaction inside the cavity of the hybrid device proliferated the absorbed energy density and effective polarizability of a nanoscale target magnetic bit surrounded by oppositely magnetized background regions in a racetrack. Numerical results demonstrated that the absorbed energy density can be enhanced up to more than 30$\times$ (depending on the width of the magnetic racetrack from 120 nm down to 30 nm), exceeding the fluence threshold of switching in addition to preventing prevalent nonlinear absorption losses in the IMOS platform. Based on the numerical results, the hybrid device can enable the detection of the magnetization state in a $\sim$200 $\times$ 100 nm\textsuperscript{2} magnetic bit with a relative contrast of $> 0.6\%$, while target bits down to $\sim$100 $\times$ 100 nm\textsuperscript{2} can be identified distinctly, no matter what the magnetization state is in the background regions in the racetrack. Nowadays, all-optical access to a spintronic platform has turned into a hot topic in the field of spintronics and we believe our proposed device can potentially bridge integrated photonics with spintronics for ultrafast and energy efficient advanced on-chip applications.\par

\section*{Methods} \label{s-method}

\noindent \textbf{Simulations of AOS and AOR.} The numerical results of our studies in AOS and AOR are conducted using commercial Lumerical software \cite{fdtd}. The power absorption analysis was carried out using the \textit{Power absorbed (advanced)} analysis group. To calculate the Kerr rotation due to PMOKE, we considered a multilayer stack of Ta(4)/Co(2)/Gd(4)/Pt(2), where the magneto-optical response is only considered for Co with a Voigt constant of $Q$ = 0.15427-0.10029$i$ \cite{van2009optimization}. Knowing that Gd shows no MO response at optical wavelengths, the refractive index of Pt is used for Gd. To measure the Kerr rotation, we exploited the \textit{Polarization ellipse} analysis group which returns the polarization rotation angle and ellipticity angle. Furthermore, perfectly matched layer (PML) absorbing boundary condition is used to define the simulation domain. The refractive indices of all the materials in the simulations are selected from the material database of Lumerical \cite{fdtd}. Finally, for accurate numerical modelling of MO activity, conformal mesh alongside an override mesh for the PNA and racetrack are defined with maximum mesh sizes of 5, 5, and 1 nm along the \textbf{X}, \textbf{Y}, and \textbf{Z} axes.\par

\vline

\noindent \textbf{Pulse propagation with high intensity in passive waveguide.} The traveling wave circuit model \enquote{PhIsim} \cite{Phisim} with an extension of nonlinear effects is used to model the propagation of the fundamental TE\textsubscript{0} component of light in an InP waveguide on the IMOS platform. The nonlinear effects are introduced through modeling the evolution of the photon density, phase, and its coupling to the charge density with the traveling distance and time. The photon density as a signature of photon power of forward ($P_1$) and backward ($P_2$) propagation directions is modeled as Eq. \ref{photonD}, where $v$\textsubscript{g} is the group velocity, $\gamma$\textsubscript{p} is the passive loss of the waveguide ($80/$m \cite{Latkowski:2015uw,Augustin:2018aa,Smit:2019vc}), $\gamma$\textsubscript{car} is the FCA coefficient ($7.2\times10^{-21}/$m \cite{Gonzalez:2009ut,Latkowski:2015uw}), $\beta$ is the TPA coefficient ($1.50\times10^{-10}$ m$/$W \cite{Gonzalez:2009ut}). $\tau$ is the carrier lifetime (2000 ps). \textit{B} and \textit{C} are respectively bi-molecular recombination constant ($2.62\times10^{-16}$ m\textsuperscript{3}$/$s \cite{Bente:2008wj}) and Auger recombination constant ($5.27\times10^{-41}$ m\textsuperscript{6}$/$s \cite{Bente:2008wj}). $\Gamma_2$ is the two photon confinement factor (0.7) as obtained from the Lumerical MODE solver \cite{mode}. $\alpha$\textsubscript{p} is the phase delay factor ($1.7\times10^{-26}$ m\textsuperscript{3} \cite{Dvorak:1995ui}). 

\begin{multline}
\frac{dP_{1,2}}{dt} = (-\gamma_p -\gamma_{car}N) v_{g} P_{1,2} \\ - 2\Gamma_2 \beta \hbar\omega v_g^2 (P_{1,2}^2+P_1P_2)
\label{photonD}
\end{multline}

\begin{equation}
\frac{d\phi_{1,2}}{dt} = \Gamma \alpha_p N - \Gamma_{2p} \frac{ \beta \hbar\omega v_g n_2}{c}(P_{1}+P_{2})
\end{equation}

\begin{equation}
\frac{dN}{dt} =  \frac{ \Gamma_2 \beta \hbar\omega v_g^2}{\Gamma }(P_{1}^2+P_{2}^2) - \frac{N}{\tau} - BN^2-CN^3
\end{equation}

\backmatter
\bmhead{Acknowledgments}
This work is part of the Gravitation programme ‘Research Centre for Integrated
Nanophotonics’, which is financed by the Netherlands Organisation for Scientific
Research (NWO). This  project  has  also received  funding  from  the  European Union’s Horizon 2020 research and innovation programme under the Marie Sklodowska-Curie grant agreement No. 860060.

\section*{Author contributions}
H.P. and P.L. conceived and designed the project. H.P. performed the designs, simulation of devices, and data analysis. P. L. performed the power/energy analysis for switching function. B.K. supervised the project. R.L., M.H., E.B., J.J.G.M.v.d.T, and B.K. contributed to the discussion and commented on the manuscript. H.P., P.L. and B.K. wrote the manuscript with input from all authors.

\section*{Competing interests}
The authors declare no competing interests.

\section*{Additional information}
Supplementary information is placed after the References section.

\bibliographystyle{naturemag}
\bibliography{MyLib}

\begin{thebibliography}{10}
\expandafter\ifx\csname url\endcsname\relax
  \def\url#1{\texttt{#1}}\fi
\expandafter\ifx\csname urlprefix\endcsname\relax\def\urlprefix{URL }\fi
\providecommand{\bibinfo}[2]{#2}
\providecommand{\eprint}[2][]{\url{#2}}

\bibitem{bogaerts2005nanophotonic}
\bibinfo{author}{Bogaerts, W.} \emph{et~al.}
\newblock \bibinfo{title}{Nanophotonic waveguides in silicon-on-insulator
  fabricated with {CMOS} technology}.
\newblock \emph{\bibinfo{journal}{J. Light. Technol.}}
  \textbf{\bibinfo{volume}{23}}, \bibinfo{pages}{401--412}
  (\bibinfo{year}{2005}).

\bibitem{bogaerts2012silicon}
\bibinfo{author}{Bogaerts, W.} \emph{et~al.}
\newblock \bibinfo{title}{Silicon microring resonators}.
\newblock \emph{\bibinfo{journal}{Laser Photonics Rev.}}
  \textbf{\bibinfo{volume}{6}}, \bibinfo{pages}{47--73} (\bibinfo{year}{2012}).

\bibitem{van2019inp}
\bibinfo{author}{van~der Tol, J.~J.} \emph{et~al.}
\newblock \bibinfo{title}{{InP} membrane on silicon ({IMOS}) photonics}.
\newblock \emph{\bibinfo{journal}{IEEE J. Quantum Electron.}}
  \textbf{\bibinfo{volume}{56}}, \bibinfo{pages}{1--7} (\bibinfo{year}{2019}).

\bibitem{jiao2020indium}
\bibinfo{author}{Jiao, Y.} \emph{et~al.}
\newblock \bibinfo{title}{Indium phosphide membrane nanophotonic integrated
  circuits on silicon}.
\newblock \emph{\bibinfo{journal}{Phys. Status Solidi A}}
  \textbf{\bibinfo{volume}{217}}, \bibinfo{pages}{1900606}
  (\bibinfo{year}{2020}).

\bibitem{Dieny:2020ub}
\bibinfo{author}{Dieny, B.} \emph{et~al.}
\newblock \bibinfo{title}{Opportunities and challenges for spintronics in the
  microelectronics industry}.
\newblock \emph{\bibinfo{journal}{Nat. Electron.}}
  \textbf{\bibinfo{volume}{3}}, \bibinfo{pages}{446--459}
  (\bibinfo{year}{2020}).

\bibitem{Parkin:2015aa}
\bibinfo{author}{Parkin, S.} \& \bibinfo{author}{Yang, S.-H.}
\newblock \bibinfo{title}{Memory on the racetrack}.
\newblock \emph{\bibinfo{journal}{Nat. Nanotechnol.}}
  \textbf{\bibinfo{volume}{10}}, \bibinfo{pages}{195--198}
  (\bibinfo{year}{2015}).

\bibitem{BlaesingECT2018}
\bibinfo{author}{Bl{\"a}sing, R.} \emph{et~al.}
\newblock \bibinfo{title}{Exchange coupling torque in ferrimagnetic {C}o/{G}d
  bilayer maximized near angular momentum compensation temperature}.
\newblock \emph{\bibinfo{journal}{Nat. Commun.}} \textbf{\bibinfo{volume}{9}},
  \bibinfo{pages}{4984} (\bibinfo{year}{2018}).

\bibitem{Li:2022aa}
\bibinfo{author}{Li, P.} \emph{et~al.}
\newblock \bibinfo{title}{Ultrafast racetrack based on compensated co/gd-based
  synthetic ferrimagnet with all-optical switching} (\bibinfo{year}{2022}).
\newblock \urlprefix\url{https://arxiv.org/abs/2204.11595}.

\bibitem{Kimel_AOS_Review2019}
\bibinfo{author}{Alexey~V.Kimel, M.~L.}
\newblock \bibinfo{title}{Writing magnetic memory with ultrashort light
  pulses}.
\newblock \emph{\bibinfo{journal}{Nat. Rev. Mater.}}
  \textbf{\bibinfo{volume}{4}}, \bibinfo{pages}{189--200}
  (\bibinfo{year}{2019}).

\bibitem{maccaferri2016anisotropic}
\bibinfo{author}{Maccaferri, N.} \emph{et~al.}
\newblock \bibinfo{title}{Anisotropic nanoantenna-based magnetoplasmonic
  crystals for highly enhanced and tunable magneto-optical activity}.
\newblock \emph{\bibinfo{journal}{Nano Lett.}} \textbf{\bibinfo{volume}{16}},
  \bibinfo{pages}{2533--2542} (\bibinfo{year}{2016}).

\bibitem{pineider2018nanomaterials}
\bibinfo{author}{Pineider, F.} \& \bibinfo{author}{Sangregorio, C.}
\newblock \bibinfo{title}{Nanomaterials for magnetoplasmonics}.
\newblock In \emph{\bibinfo{booktitle}{Novel Magnetic Nanostructures}},
  \bibinfo{pages}{191--220} (\bibinfo{publisher}{Elsevier},
  \bibinfo{year}{2018}).

\bibitem{demirer2021magneto}
\bibinfo{author}{Demirer, F.~E.} \emph{et~al.}
\newblock \bibinfo{title}{Magneto-photonic on-chip device for all-optical
  reading of magnetic memory}.
\newblock In \emph{\bibinfo{booktitle}{2021 Conference on Lasers and
  Electro-Optics Europe \& European Quantum Electronics Conference
  (CLEO/Europe-EQEC)}}, \bibinfo{pages}{1--1} (\bibinfo{organization}{IEEE},
  \bibinfo{year}{2021}).

\bibitem{demirer2022integrated}
\bibinfo{author}{Demirer, F.~E.} \emph{et~al.}
\newblock \bibinfo{title}{An integrated photonic device for on-chip
  magneto-optical memory reading}.
\newblock \emph{\bibinfo{journal}{Nanophotonics}}
  \textbf{\bibinfo{volume}{11}}, \bibinfo{pages}{3319--3329}
  (\bibinfo{year}{2022}).

\bibitem{hamedAOR}
\bibinfo{author}{Pezeshki, H.} \emph{et~al.}
\newblock \bibinfo{title}{Optical reading of nanoscale magnetic bits in an
  integrated photonic platform} (\bibinfo{year}{2022}).
\newblock \urlprefix\url{https://arxiv.org/abs/2208.02560}.

\bibitem{Ostler:2012aa}
\bibinfo{author}{Ostler, T.~A.} \emph{et~al.}
\newblock \bibinfo{title}{Ultrafast heating as a sufficient stimulus for
  magnetization reversal in a ferrimagnet}.
\newblock \emph{\bibinfo{journal}{Nat. Commun.}} \textbf{\bibinfo{volume}{3}},
  \bibinfo{pages}{666} (\bibinfo{year}{2012}).

\bibitem{Radu:2011aa}
\bibinfo{author}{Radu, I.} \emph{et~al.}
\newblock \bibinfo{title}{Transient ferromagnetic-like state mediating
  ultrafast reversal of antiferromagnetically coupled spins}.
\newblock \emph{\bibinfo{journal}{Nature}} \textbf{\bibinfo{volume}{472}},
  \bibinfo{pages}{205--208} (\bibinfo{year}{2011}).

\bibitem{Stanciu:2007aa}
\bibinfo{author}{Stanciu, C.~D.} \emph{et~al.}
\newblock \bibinfo{title}{All-optical magnetic recording with circularly
  polarized light}.
\newblock \emph{\bibinfo{journal}{Phys. Rev. Lett.}}
  \textbf{\bibinfo{volume}{99}}, \bibinfo{pages}{047601--}
  (\bibinfo{year}{2007}).

\bibitem{Lalieu:2017aa}
\bibinfo{author}{Lalieu, M. L.~M.} \emph{et~al.}
\newblock \bibinfo{title}{Deterministic all-optical switching of synthetic
  ferrimagnets using single femtosecond laser pulses}.
\newblock \emph{\bibinfo{journal}{Phys. Rev. B}} \textbf{\bibinfo{volume}{96}},
  \bibinfo{pages}{220411--} (\bibinfo{year}{2017}).

\bibitem{Aviles-Felix:2019aa}
\bibinfo{author}{Avil{\'e}s-F{\'e}lix, L.} \emph{et~al.}
\newblock \bibinfo{title}{Integration of {Tb/Co} multilayers within optically
  switchable perpendicular magnetic tunnel junctions}.
\newblock \emph{\bibinfo{journal}{AIP Adv.}} \textbf{\bibinfo{volume}{9}},
  \bibinfo{pages}{125328} (\bibinfo{year}{2019}).

\bibitem{Aviles-Felix:2020aa}
\bibinfo{author}{Avil{\'e}s-F{\'e}lix, L.} \emph{et~al.}
\newblock \bibinfo{title}{Single-shot all-optical switching of magnetization in
  {Tb/Co} multilayer-based electrodes}.
\newblock \emph{\bibinfo{journal}{Sci. Rep.}} \textbf{\bibinfo{volume}{10}},
  \bibinfo{pages}{5211} (\bibinfo{year}{2020}).

\bibitem{Li:2021wr}
\bibinfo{author}{Li, P.} \emph{et~al.}
\newblock \bibinfo{title}{Ultra-low energy threshold engineering for
  all-optical switching of magnetization in dielectric-coated {Co/Gd} based
  synthetic-ferrimagnet}.
\newblock \emph{\bibinfo{journal}{Appl. Phys. Lett.}}
  \textbf{\bibinfo{volume}{119}}, \bibinfo{pages}{252402}
  (\bibinfo{year}{2021}).

\bibitem{Luding:2022ur}
\bibinfo{author}{Wang, L.} \emph{et~al.}
\newblock \bibinfo{title}{Picosecond optospintronic tunnel junctions}.
\newblock \emph{\bibinfo{journal}{Proc. Natl. Acad. Sci.}}
  \textbf{\bibinfo{volume}{119}}, \bibinfo{pages}{e2204732119}
  (\bibinfo{year}{2022}).

\bibitem{Wang:2020ab}
\bibinfo{author}{Wang, L.} \emph{et~al.}
\newblock \bibinfo{title}{Enhanced all-optical switching and domain wall
  velocity in annealed synthetic-ferrimagnetic multilayers}.
\newblock \emph{\bibinfo{journal}{Appl. Phys. Lett.}}
  \textbf{\bibinfo{volume}{117}}, \bibinfo{pages}{022408}
  (\bibinfo{year}{2020}).

\bibitem{Sobolewska:2020aa}
\bibinfo{author}{Sobolewska, E.~K.} \emph{et~al.}
\newblock \bibinfo{title}{Integration platform for optical switching of
  magnetic elements}.
\newblock In \emph{\bibinfo{booktitle}{Proc.SPIE}}, vol.
  \bibinfo{volume}{11461} (\bibinfo{year}{2020}).

\bibitem{Chen:2017aa}
\bibinfo{author}{Chen, J.-Y.} \emph{et~al.}
\newblock \bibinfo{title}{All-optical switching of magnetic tunnel junctions
  with single subpicosecond laser pulses}.
\newblock \emph{\bibinfo{journal}{Phys. Rev. Appl.}}
  \textbf{\bibinfo{volume}{7}}, \bibinfo{pages}{021001--}
  (\bibinfo{year}{2017}).

\bibitem{hutter2004exploitation}
\bibinfo{author}{Hutter, E.} \& \bibinfo{author}{Fendler, J.~H.}
\newblock \bibinfo{title}{Exploitation of localized surface plasmon resonance}.
\newblock \emph{\bibinfo{journal}{Adv. Mater.}} \textbf{\bibinfo{volume}{16}},
  \bibinfo{pages}{1685--1706} (\bibinfo{year}{2004}).

\bibitem{gramotnev2010plasmonics}
\bibinfo{author}{Gramotnev, D.~K.} \& \bibinfo{author}{Bozhevolnyi, S.~I.}
\newblock \bibinfo{title}{Plasmonics beyond the diffraction limit}.
\newblock \emph{\bibinfo{journal}{Nat. Photon.}} \textbf{\bibinfo{volume}{4}},
  \bibinfo{pages}{83--91} (\bibinfo{year}{2010}).

\bibitem{pezeshki2021ultra}
\bibinfo{author}{Pezeshki, H.}, \bibinfo{author}{Wright, A.~J.} \&
  \bibinfo{author}{Larkins, E.~C.}
\newblock \bibinfo{title}{Ultra-compact and ultra-broadband hybrid
  plasmonic-photonic vertical coupler with high coupling efficiency,
  directivity, and polarisation extinction ratio}.
\newblock \emph{\bibinfo{journal}{IET Optoelectron.}}  (\bibinfo{year}{2021}).

\bibitem{Pezeshki:22}
\bibinfo{author}{Pezeshki, H.}
\newblock \bibinfo{title}{Highly efficient, ultra-compact, and ultra-broadband
  bidirectional vertical coupler based on spin-directional locking}.
\newblock \emph{\bibinfo{journal}{J. Opt. Soc. Am. B}}
  \textbf{\bibinfo{volume}{39}}, \bibinfo{pages}{2714--2722}
  (\bibinfo{year}{2022}).

\bibitem{mejia2018plasmonic}
\bibinfo{author}{Mej{\'\i}a-Salazar, J.} \& \bibinfo{author}{Oliveira~Jr,
  O.~N.}
\newblock \bibinfo{title}{Plasmonic biosensing: Focus review}.
\newblock \emph{\bibinfo{journal}{Chem. Rev.}} \textbf{\bibinfo{volume}{118}},
  \bibinfo{pages}{10617--10625} (\bibinfo{year}{2018}).

\bibitem{pezeshki2021lab}
\bibinfo{author}{Pezeshki, H.}
\newblock \emph{\bibinfo{title}{Lab-on-a-chip technology platform for
  biophotonic applications}}.
\newblock Ph.D. thesis, \bibinfo{school}{University of Nottingham}
  (\bibinfo{year}{2021}).

\bibitem{zain2007tapered}
\bibinfo{author}{Zain, A. R.~M.} \emph{et~al.}
\newblock \bibinfo{title}{Tapered photonic crystal microcavities embedded in
  photonic wire waveguides with large resonance quality-factor and high
  transmission}.
\newblock \emph{\bibinfo{journal}{IEEE Photonics Technol. Lett.}}
  \textbf{\bibinfo{volume}{20}}, \bibinfo{pages}{6--8} (\bibinfo{year}{2007}).

\bibitem{pezeshki2015design}
\bibinfo{author}{Pezeshki, H.} \& \bibinfo{author}{Darvish, G.}
\newblock \bibinfo{title}{Design of photonic crystal microcavity based optical
  switches using fano resonance effect}.
\newblock \emph{\bibinfo{journal}{Optik}} \textbf{\bibinfo{volume}{126}},
  \bibinfo{pages}{4202--4205} (\bibinfo{year}{2015}).

\bibitem{reniers2019characterization}
\bibinfo{author}{Reniers, S.~F.} \emph{et~al.}
\newblock \bibinfo{title}{Characterization of waveguide photonic crystal
  reflectors on indium phosphide membranes}.
\newblock \emph{\bibinfo{journal}{IEEE J. Quantum Electron.}}
  \textbf{\bibinfo{volume}{55}}, \bibinfo{pages}{1--7} (\bibinfo{year}{2019}).

\bibitem{yablonovitch1994photonic}
\bibinfo{author}{Yablonovitch, E.}
\newblock \bibinfo{title}{Photonic crystals}.
\newblock \emph{\bibinfo{journal}{J. Mod. Opt.}} \textbf{\bibinfo{volume}{41}},
  \bibinfo{pages}{173--194} (\bibinfo{year}{1994}).

\bibitem{joannopoulos1997photonic}
\bibinfo{author}{Joannopoulos, J.~D.}, \bibinfo{author}{Villeneuve, P.~R.} \&
  \bibinfo{author}{Fan, S.}
\newblock \bibinfo{title}{Photonic crystals: putting a new twist on light}.
\newblock \emph{\bibinfo{journal}{Nature}} \textbf{\bibinfo{volume}{386}},
  \bibinfo{pages}{143--149} (\bibinfo{year}{1997}).

\bibitem{ellis2011ultralow}
\bibinfo{author}{Ellis, B.} \emph{et~al.}
\newblock \bibinfo{title}{Ultralow-threshold electrically pumped quantum-dot
  photonic-crystal nanocavity laser}.
\newblock \emph{\bibinfo{journal}{Nat. Photon.}} \textbf{\bibinfo{volume}{5}},
  \bibinfo{pages}{297--300} (\bibinfo{year}{2011}).

\bibitem{pezeshki2013all}
\bibinfo{author}{Pezeshki, H.} \& \bibinfo{author}{Ahmadi, V.}
\newblock \bibinfo{title}{All-optical bistable switching based on photonic
  crystal slab nanocavity using nonlinear kerr effect}.
\newblock \emph{\bibinfo{journal}{J. Mod. Opt.}} \textbf{\bibinfo{volume}{60}},
  \bibinfo{pages}{103--108} (\bibinfo{year}{2013}).

\bibitem{pezeshki2016design}
\bibinfo{author}{Pezeshki, H.} \& \bibinfo{author}{Darvish, G.}
\newblock \bibinfo{title}{Design and simulation of photonic crystal based
  all-optical logic gate and modulator using infiltration}.
\newblock \emph{\bibinfo{journal}{Opt. Quantum Electron.}}
  \textbf{\bibinfo{volume}{48}}, \bibinfo{pages}{1--15} (\bibinfo{year}{2016}).

\bibitem{Parkin:2008aa}
\bibinfo{author}{Parkin, S. S.~P.}, \bibinfo{author}{Hayashi, M.} \&
  \bibinfo{author}{Thomas, L.}
\newblock \bibinfo{title}{Magnetic domain-wall racetrack memory}.
\newblock \emph{\bibinfo{journal}{Science}} \textbf{\bibinfo{volume}{320}},
  \bibinfo{pages}{190} (\bibinfo{year}{2008}).

\bibitem{blasing2020magnetic}
\bibinfo{author}{Bl{\"a}sing, R.} \emph{et~al.}
\newblock \bibinfo{title}{Magnetic racetrack memory: From physics to the cusp
  of applications within a decade}.
\newblock \emph{\bibinfo{journal}{Proc. IEEE}} \textbf{\bibinfo{volume}{108}},
  \bibinfo{pages}{1303--1321} (\bibinfo{year}{2020}).

\bibitem{li2021ultra}
\bibinfo{author}{Li, P.} \emph{et~al.}
\newblock \bibinfo{title}{Ultra-low energy threshold engineering for
  all-optical switching of magnetization in dielectric-coated {Co/Gd} based
  synthetic-ferrimagnet}.
\newblock \emph{\bibinfo{journal}{Appl. Phys. Lett.}}
  \textbf{\bibinfo{volume}{119}}, \bibinfo{pages}{252402}
  (\bibinfo{year}{2021}).

\bibitem{Miron:2011aa}
\bibinfo{author}{Miron, I.~M.} \emph{et~al.}
\newblock \bibinfo{title}{Perpendicular switching of a single ferromagnetic
  layer induced by in-plane current injection}.
\newblock \emph{\bibinfo{journal}{Nature}} \textbf{\bibinfo{volume}{476}},
  \bibinfo{pages}{189--193} (\bibinfo{year}{2011}).

\bibitem{Ryu:2013aa}
\bibinfo{author}{Ryu, K.-S.} \emph{et~al.}
\newblock \bibinfo{title}{Chiral spin torque at magnetic domain walls}.
\newblock \emph{\bibinfo{journal}{Nat. Nanotechnol.}}
  \textbf{\bibinfo{volume}{8}}, \bibinfo{pages}{527--533}
  (\bibinfo{year}{2013}).

\bibitem{fdtd}
\bibinfo{title}{Lumerical inc}.
\newblock \bibinfo{howpublished}{https://www.lumerical.com/products/fdtd/}
  (\bibinfo{year}{2021}).

\bibitem{subramanian2013low}
\bibinfo{author}{Subramanian, A.} \emph{et~al.}
\newblock \bibinfo{title}{Low-loss singlemode {PECVD} silicon nitride photonic
  wire waveguides for 532--900 nm wavelength window fabricated within a {CMOS}
  pilot line}.
\newblock \emph{\bibinfo{journal}{IEEE Photon. J.}}
  \textbf{\bibinfo{volume}{5}}, \bibinfo{pages}{2202809--2202809}
  (\bibinfo{year}{2013}).

\bibitem{Mentink:2012aa}
\bibinfo{author}{Mentink, J.~H.} \emph{et~al.}
\newblock \bibinfo{title}{Ultrafast spin dynamics in multisublattice magnets}.
\newblock \emph{\bibinfo{journal}{Phys. Rev. Lett.}}
  \textbf{\bibinfo{volume}{108}}, \bibinfo{pages}{057202--}
  (\bibinfo{year}{2012}).

\bibitem{Beens:2019aa}
\bibinfo{author}{Beens, M.} \emph{et~al.}
\newblock \bibinfo{title}{Comparing all-optical switching in
  synthetic-ferrimagnetic multilayers and alloys}.
\newblock \emph{\bibinfo{journal}{Phys. Rev. B}}
  \textbf{\bibinfo{volume}{100}}, \bibinfo{pages}{220409--}
  (\bibinfo{year}{2019}).

\bibitem{Gerlach:2017aa}
\bibinfo{author}{Gerlach, S.} \emph{et~al.}
\newblock \bibinfo{title}{Modeling ultrafast all-optical switching in synthetic
  ferrimagnets}.
\newblock \emph{\bibinfo{journal}{Phys. Rev. B}} \textbf{\bibinfo{volume}{95}},
  \bibinfo{pages}{224435--} (\bibinfo{year}{2017}).

\bibitem{Latkowski:2015uw}
\bibinfo{author}{Latkowski, S.} \emph{et~al.}
\newblock \bibinfo{title}{Monolithically integrated 2.5 {GHz} extended cavity
  mode-locked ring laser with intracavity phase modulators}.
\newblock \emph{\bibinfo{journal}{Opt. Lett.}} \textbf{\bibinfo{volume}{40}},
  \bibinfo{pages}{77--80} (\bibinfo{year}{2015}).

\bibitem{Thourhout:2001vm}
\bibinfo{author}{van Thourhout, D.} \emph{et~al.}
\newblock \bibinfo{title}{Observation of {WDM} crosstalk in passive
  semiconductor waveguides}.
\newblock \emph{\bibinfo{journal}{IEEE Photonics Technol. Lett.}}
  \textbf{\bibinfo{volume}{13}}, \bibinfo{pages}{457--459}
  (\bibinfo{year}{2001}).

\bibitem{Gonzalez:2009ut}
\bibinfo{author}{Gonzalez, L.~P.} \emph{et~al.}
\newblock \bibinfo{title}{Wavelength dependence of two photon and free carrier
  absorptions in inp}.
\newblock \emph{\bibinfo{journal}{Opt. Express}} \textbf{\bibinfo{volume}{17}},
  \bibinfo{pages}{8741--8748} (\bibinfo{year}{2009}).

\bibitem{Gorchon:2016aa}
\bibinfo{author}{Gorchon, J.} \emph{et~al.}
\newblock \bibinfo{title}{Role of electron and phonon temperatures in the
  helicity-independent all-optical switching of {GdFeCo}}.
\newblock \emph{\bibinfo{journal}{Phys. Rev. B}} \textbf{\bibinfo{volume}{94}},
  \bibinfo{pages}{184406--} (\bibinfo{year}{2016}).

\bibitem{Li2022He}
\bibinfo{author}{Li, P.} \emph{et~al.}
\newblock \bibinfo{title}{Enhancing all-optical switching of magnetization by
  he ion irradiation} (\bibinfo{year}{2022}).
\newblock \urlprefix\url{https://arxiv.org/abs/2207.07766}.

\bibitem{Saleh:2007aa}
\bibinfo{author}{Saleh, B. E.~A.} \& \bibinfo{author}{Teich, M.~C.}
\newblock \emph{\bibinfo{title}{Fundamentals of Photonics}}
  (\bibinfo{publisher}{Wiley}, \bibinfo{year}{2007}).

\bibitem{Wei:2021ui}
\bibinfo{author}{Wei, J.} \emph{et~al.}
\newblock \bibinfo{title}{All-optical helicity-independent switching state
  diagram in {Gd-Fe-Co} alloys}.
\newblock \emph{\bibinfo{journal}{Phys. Rev. Appl.}}
  \textbf{\bibinfo{volume}{15}}, \bibinfo{pages}{054065}
  (\bibinfo{year}{2021}).

\bibitem{Phisim}
\bibinfo{author}{Bente, E.}
\newblock \bibinfo{title}{Phisim - a photonic integrated circuit simulator}.
\newblock \bibinfo{howpublished}{https://sites.google.com/tue.nl/phisim/home}
  (\bibinfo{year}{2021}).

\bibitem{Gil-Molina:2018us}
\bibinfo{author}{Gil-Molina, A.} \emph{et~al.}
\newblock \bibinfo{title}{Optical free-carrier generation in silicon
  nano-waveguides at 1550 nm}.
\newblock \emph{\bibinfo{journal}{Appl. Phys. Lett.}}
  \textbf{\bibinfo{volume}{112}}, \bibinfo{pages}{251104}
  (\bibinfo{year}{2018}).

\bibitem{Black:2017ub}
\bibinfo{author}{Black, L.~E.} \emph{et~al.}
\newblock \bibinfo{title}{Effective surface passivation of inp nanowires by
  atomic-layer-deposited al2o3 with pox interlayer}.
\newblock \emph{\bibinfo{journal}{Nano Lett.}} \textbf{\bibinfo{volume}{17}},
  \bibinfo{pages}{6287--6294} (\bibinfo{year}{2017}).

\bibitem{Bente:2021th}
\bibinfo{author}{Bente, E.} \emph{et~al.}
\newblock \bibinfo{title}{Effects of two-photon absorption and non-linear index
  in inp-based passive waveguides on integrated extended cavity semiconductor
  lasers}.
\newblock In \emph{\bibinfo{booktitle}{2021 Conference on Lasers and
  Electro-Optics Europe \& European Quantum Electronics Conference
  (CLEO/Europe-EQEC)}}, \bibinfo{pages}{1--1} (\bibinfo{organization}{IEEE},
  \bibinfo{year}{2021}).

\bibitem{mode}
\bibinfo{title}{Lumerical inc}.
\newblock \bibinfo{howpublished}{https://www.lumerical.com/products/mode/}
  (\bibinfo{year}{2021}).

\bibitem{kataja2015surface}
\bibinfo{author}{Kataja, M.} \emph{et~al.}
\newblock \bibinfo{title}{Surface lattice resonances and magneto-optical
  response in magnetic nanoparticle arrays}.
\newblock \emph{\bibinfo{journal}{Nat. Commun.}} \textbf{\bibinfo{volume}{6}},
  \bibinfo{pages}{1--8} (\bibinfo{year}{2015}).

\bibitem{kravets2018plasmonic}
\bibinfo{author}{Kravets, V.~G.} \emph{et~al.}
\newblock \bibinfo{title}{Plasmonic surface lattice resonances: a review of
  properties and applications}.
\newblock \emph{\bibinfo{journal}{Chem. Rev.}} \textbf{\bibinfo{volume}{118}},
  \bibinfo{pages}{5912--5951} (\bibinfo{year}{2018}).

\bibitem{johnson1972optical}
\bibinfo{author}{Johnson, P.~B.} \& \bibinfo{author}{Christy, R.-W.}
\newblock \bibinfo{title}{Optical constants of the noble metals}.
\newblock \emph{\bibinfo{journal}{Phys. Rev. B}} \textbf{\bibinfo{volume}{6}},
  \bibinfo{pages}{4370} (\bibinfo{year}{1972}).

\bibitem{Garello:2019aa}
\bibinfo{author}{Garello, K.} \emph{et~al.}
\newblock \bibinfo{title}{Manufacturable 300mm platform solution for field-free
  switching sot-mram}.
\newblock In \emph{\bibinfo{booktitle}{2019 Symposium on VLSI Circuits}},
  \bibinfo{pages}{T194--T195} (\bibinfo{year}{2019}).

\bibitem{gao2015chip}
\bibinfo{author}{Gao, L.} \emph{et~al.}
\newblock \bibinfo{title}{On-chip plasmonic waveguide optical waveplate}.
\newblock \emph{\bibinfo{journal}{Sci. Rep.}} \textbf{\bibinfo{volume}{5}},
  \bibinfo{pages}{1--6} (\bibinfo{year}{2015}).

\bibitem{deng2005design}
\bibinfo{author}{Deng, H.} \emph{et~al.}
\newblock \bibinfo{title}{Design rules for slanted-angle polarization
  rotators}.
\newblock \emph{\bibinfo{journal}{J. Light. Technol.}}
  \textbf{\bibinfo{volume}{23}}, \bibinfo{pages}{432} (\bibinfo{year}{2005}).

\bibitem{gao2013ultra}
\bibinfo{author}{Gao, L.} \emph{et~al.}
\newblock \bibinfo{title}{Ultra-compact and low-loss polarization rotator based
  on asymmetric hybrid plasmonic waveguide}.
\newblock \emph{\bibinfo{journal}{IEEE Photonics Technol. Lett.}}
  \textbf{\bibinfo{volume}{25}}, \bibinfo{pages}{2081--2084}
  (\bibinfo{year}{2013}).

\bibitem{collett2005field}
\bibinfo{author}{Collett, E.}
\newblock \bibinfo{title}{Field guide to polarization}
  (\bibinfo{organization}{Spie Bellingham, WA}, \bibinfo{year}{2005}).

\bibitem{Becker:2020aa}
\bibinfo{author}{Becker, H.} \emph{et~al.}
\newblock \bibinfo{title}{Out-of-plane focusing grating couplers for silicon
  photonics integration with optical mram technology}.
\newblock \emph{\bibinfo{journal}{IEEE J. Sel. Top. Quantum Electron.}}
  \textbf{\bibinfo{volume}{26}}, \bibinfo{pages}{1--8} (\bibinfo{year}{2020}).

\bibitem{van2009optimization}
\bibinfo{author}{Van~Parys, W.}
\newblock \emph{\bibinfo{title}{Optimization of an Integrated Optical Isolator
  Based on a Semiconductor Amplifier with a Ferromagnetic Metal Contact.}}
\newblock Ph.D. thesis, \bibinfo{school}{Ghent University}
  (\bibinfo{year}{2009}).

\bibitem{Augustin:2018aa}
\bibinfo{author}{Augustin, L.~M.} \emph{et~al.}
\newblock \bibinfo{title}{Inp-based generic foundry platform for photonic
  integrated circuits}.
\newblock \emph{\bibinfo{journal}{IEEE J. Sel. Top. Quantum Electron.}}
  \textbf{\bibinfo{volume}{24}}, \bibinfo{pages}{1--10} (\bibinfo{year}{2018}).

\bibitem{Smit:2019vc}
\bibinfo{author}{Smit, M.}, \bibinfo{author}{Williams, K.} \&
  \bibinfo{author}{van~der Tol, J.}
\newblock \bibinfo{title}{Past, present, and future of {InP-based} photonic
  integration}.
\newblock \emph{\bibinfo{journal}{APL Photonics}} \textbf{\bibinfo{volume}{4}},
  \bibinfo{pages}{050901} (\bibinfo{year}{2019}).

\bibitem{Bente:2008wj}
\bibinfo{author}{Bente, E. A. J.~M.} \emph{et~al.}
\newblock \bibinfo{title}{Modeling of integrated extended cavity {InP/InGaAsP}
  semiconductor modelocked ring lasers}.
\newblock \emph{\bibinfo{journal}{Opt. Quantum Electron.}}
  \textbf{\bibinfo{volume}{40}}, \bibinfo{pages}{131--148}
  (\bibinfo{year}{2008}).

\bibitem{Dvorak:1995ui}
\bibinfo{author}{Dvorak, M.~D.} \emph{et~al.}
\newblock \bibinfo{title}{Nonlinear absorption and refraction of quantum
  confined {InP} nanocrystals grown in porous glass}.
\newblock \emph{\bibinfo{journal}{Appl. Phys. Lett.}}
  \textbf{\bibinfo{volume}{66}}, \bibinfo{pages}{804--806}
  (\bibinfo{year}{1995}).

\end{thebibliography}



\title{\large \textbf{Supplementary information:}\protect\\ Design of an integrated hybrid plasmonic-photonic device for all-optical switching and reading of spintronic memory}

\section*{SUPPLEMENTARY NOTE 1: PMOKE consideration in designing the waveguide} \label{s-Waveguide}

            \begin{figure*}[b!]
             \centering
                \begin{subfigure}[t]{0.49\textwidth}
                \centering
                 \includegraphics[scale = 0.25]{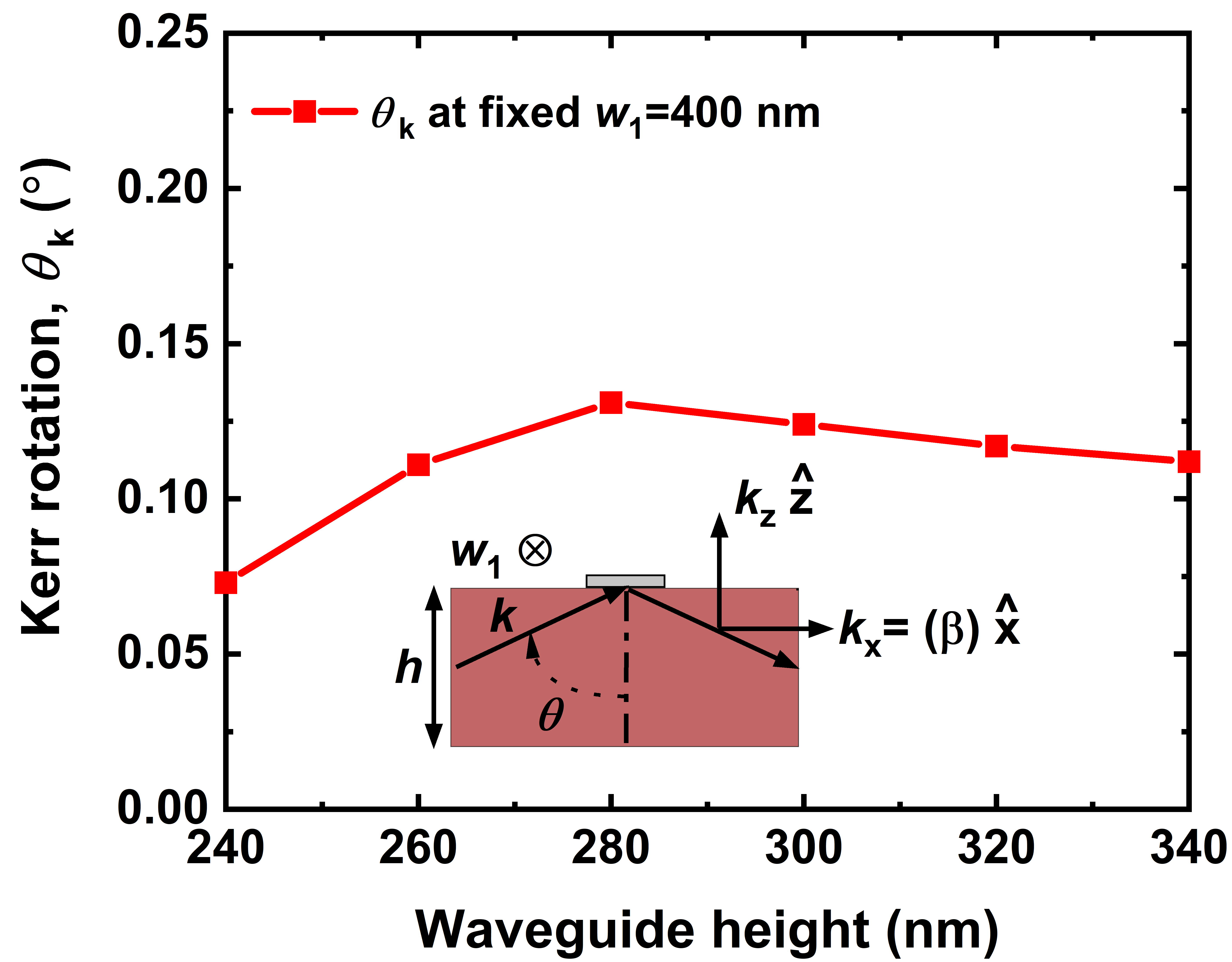}
                 \caption{}
                   \label{fig 1a}
                 \end{subfigure}
                 \begin{subfigure}[t]{0.49\textwidth}
                \centering
                \includegraphics[scale = 0.25]{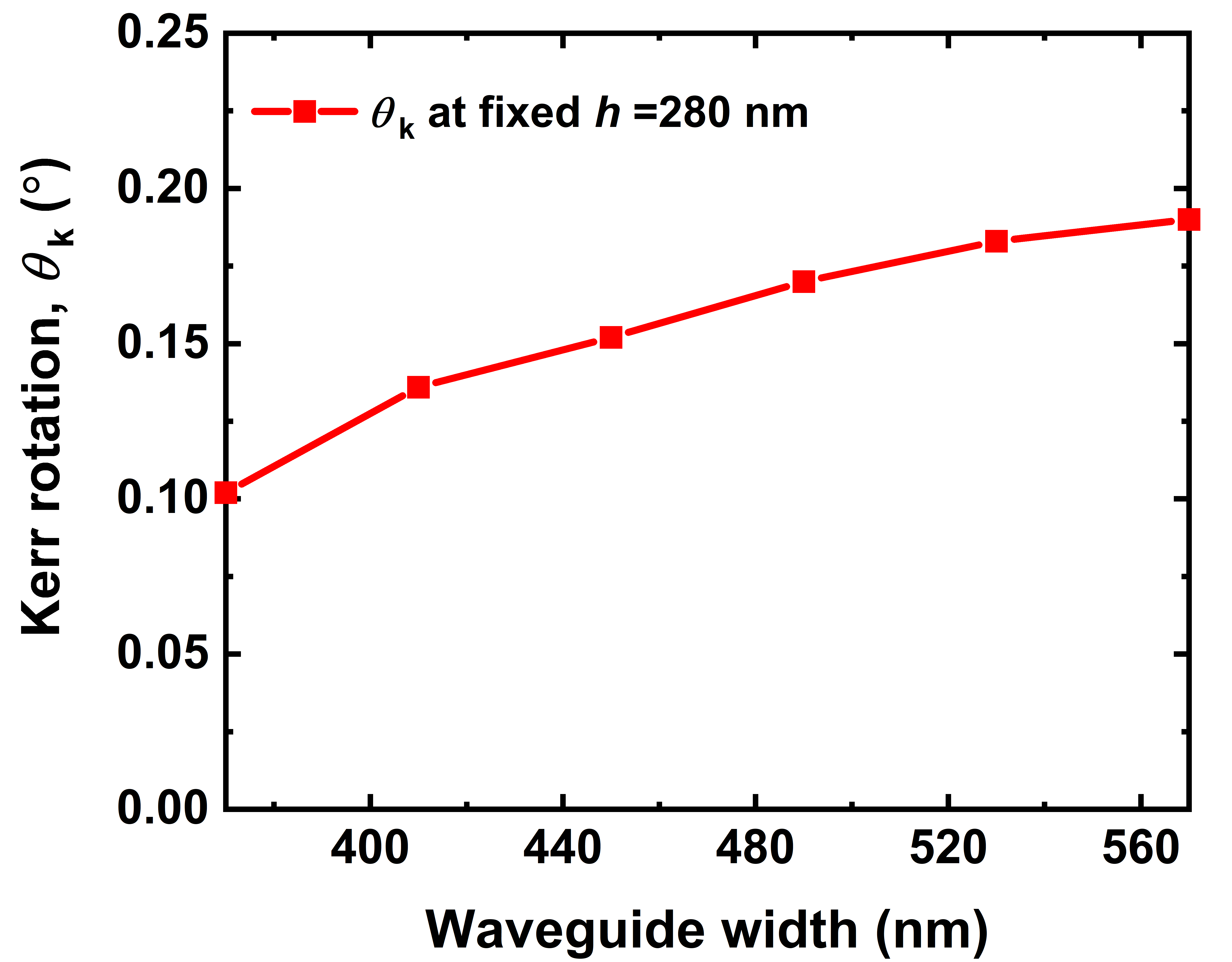}
                \caption{}
                 \label{fig 1b}
                \end{subfigure}
                 \centering
                 \caption{\small Polar magneto-optical Kerr effect (PMOKE) consideration for all optical reading. (a, b) The Kerr rotation, $\theta$\textsubscript{K}, in terms of the height, $h$, and width, $w_1$, of the waveguide, respectively.}
                \label{fig 1}
                \end{figure*}

In order to enhance the Kerr rotation, i.e. $\theta$\textsubscript{K}, the waveguide height and width are optimized. Fig. \ref{fig 1} shows $\theta$\textsubscript{K} in terms of the width ($w_1$) and height ($h$) of the waveguide, where $w_1$ is normal to the \textbf{XZ} plane as indicated in the inset. Here, $\theta$\textsubscript{K} is obtained for the same magnetic stack as the one used for the racetrack, but with a width and a length of 400$\times$300 nm\textsuperscript{2}. Figure \ref{fig 1a} shows that for an initial waveguide width of $w_1$ = 400 nm, we have a maximum $\theta$\textsubscript{K} for $h$ = 280 nm. Note that the decrease in the $\theta$\textsubscript{K} for h $<$ 280 nm is due to the fact that the waveguide can no longer guide the transverse magnetic (TM\textsubscript{0}) mode through the waveguide. By choosing $h$ = 280 nm, we inspected the impact of the waveguide's width on $\theta$\textsubscript{K}, where based on the results, we can see that further increasing of $w_1$ enhances $\theta$\textsubscript{K}, due to the larger confinement of the transverse electric (TE\textsubscript{0}) mode through the waveguide. However, we chose $w_1$ = 570 nm to avoid the excitation of higher order modes, in particular TE\textsubscript{1}.\par

\section*{SUPPLEMENTARY NOTE 2: Design of photonic crystal cavity} \label{s-phc} 

According to our application requirements, i.e. all-optical switching and reading of a spintronic memory, we designed a photonic crystal (PhC) cavity to have a defect mode resonating at $\lambda_0$ = 1.55 $\mu$m. In addition, in order to avoid pulse broadening, which can have a negative impact on the switching performance, we designed the PhC cavity with a low quality factor. Figure \ref{fig 3a} depicts a schematic of the PhC cavity, where the diameters of the holes are defined as \textit{d}\textsubscript{1-4} = 100, 110, 120, and 170 nm, and the pitch is indicated by \textit{p} = 370 nm. As shown in Fig. \ref{fig 3b}, the PhC cavity has a resonance with full width at half maximum (FWHM) of 44 nm that guarantees a low quality factor. The two dimensional electric field distribution through the waveguide (Fig. \ref{fig 2c}) illustrates the emergence of the resonance PhC cavity mode at $\lambda_0$ = 1.55 $\mu$m with an electric field enhancement of $>$ 2.5$\times$.\par 
\vspace{1cm}

            \begin{figure*}[b!]
             \centering
                \begin{subfigure}[h!]{\textwidth}
                \centering
                 \includegraphics[scale = 0.1]{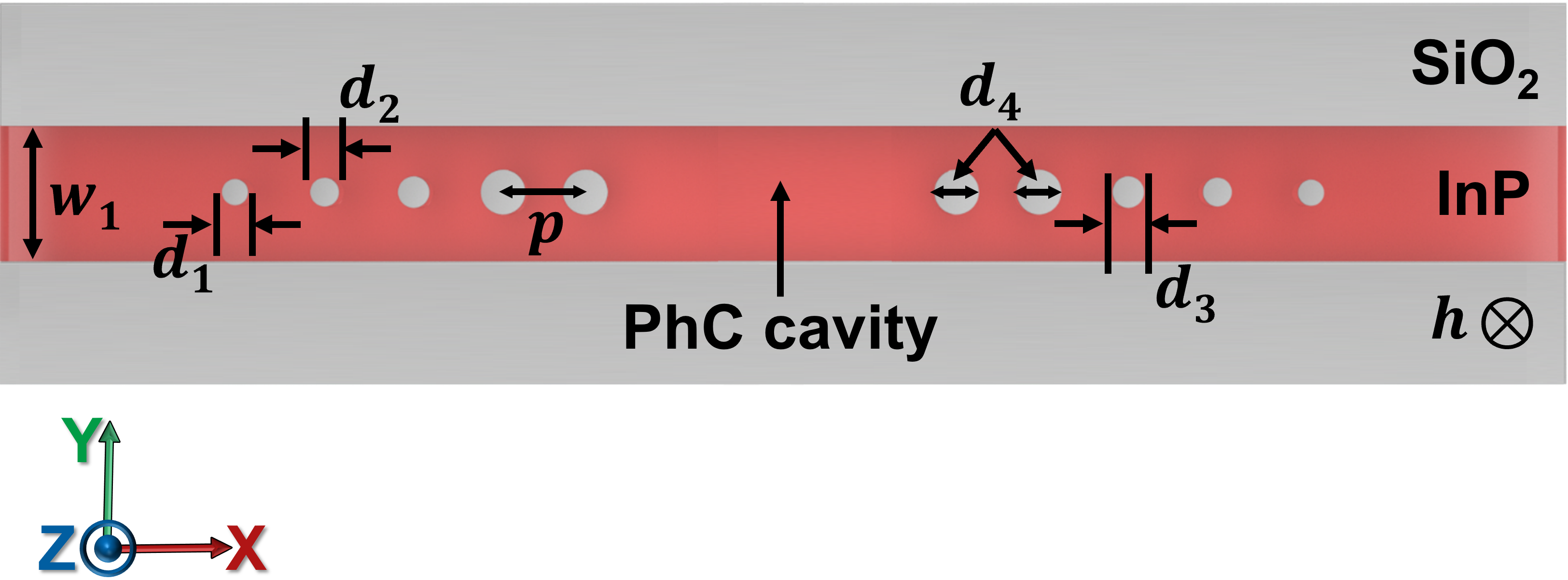}
                 \caption{}
                   \label{fig 3a}
                 \end{subfigure}
                 \begin{subfigure}[t]{0.49\textwidth}
                \centering
                \includegraphics[scale = 0.25]{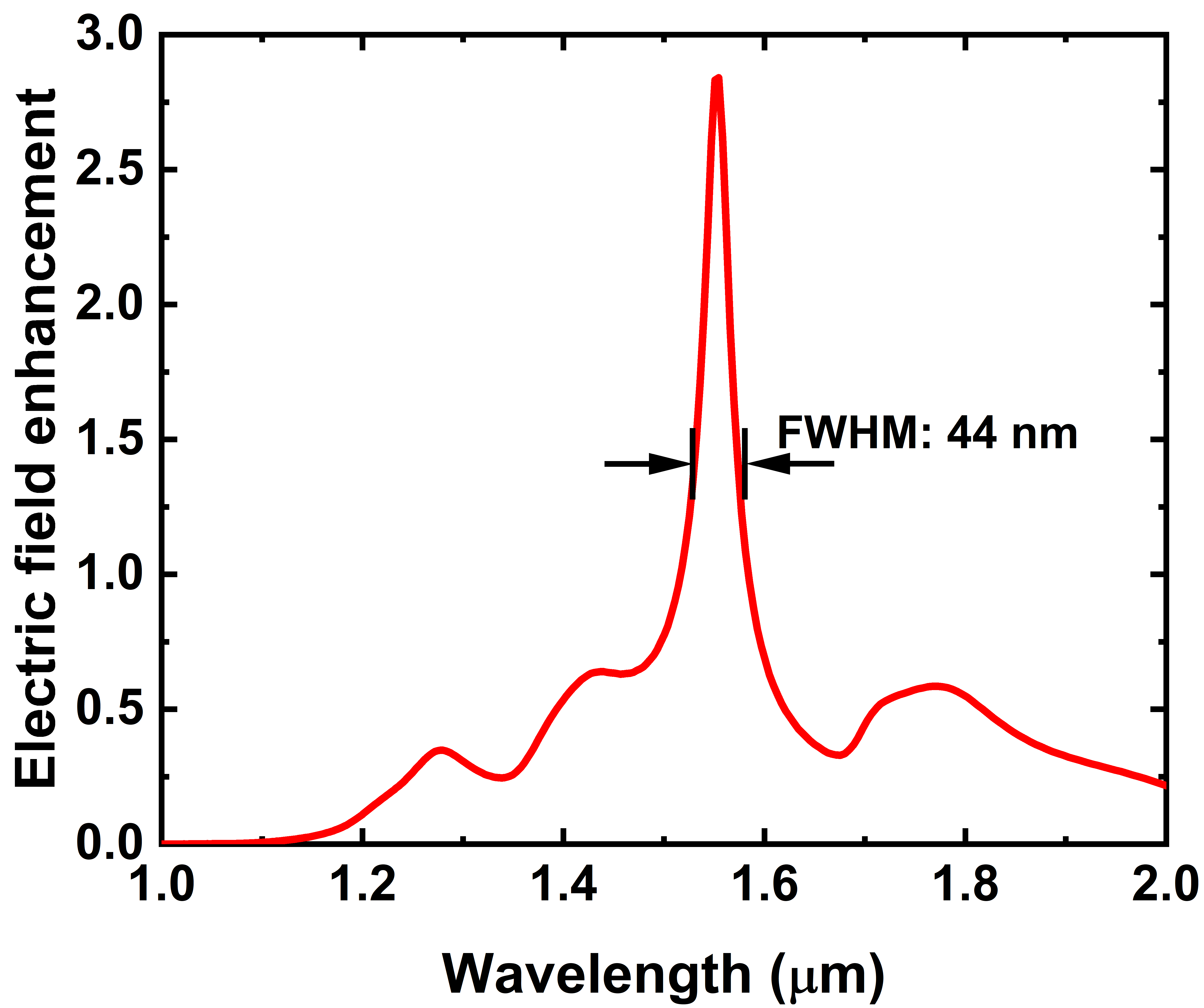}
                \caption{}
                 \label{fig 3b}
                \end{subfigure}
                 \centering
                 \begin{subfigure}[t]{0.49\textwidth}
                \centering
                \includegraphics[scale = 0.25]{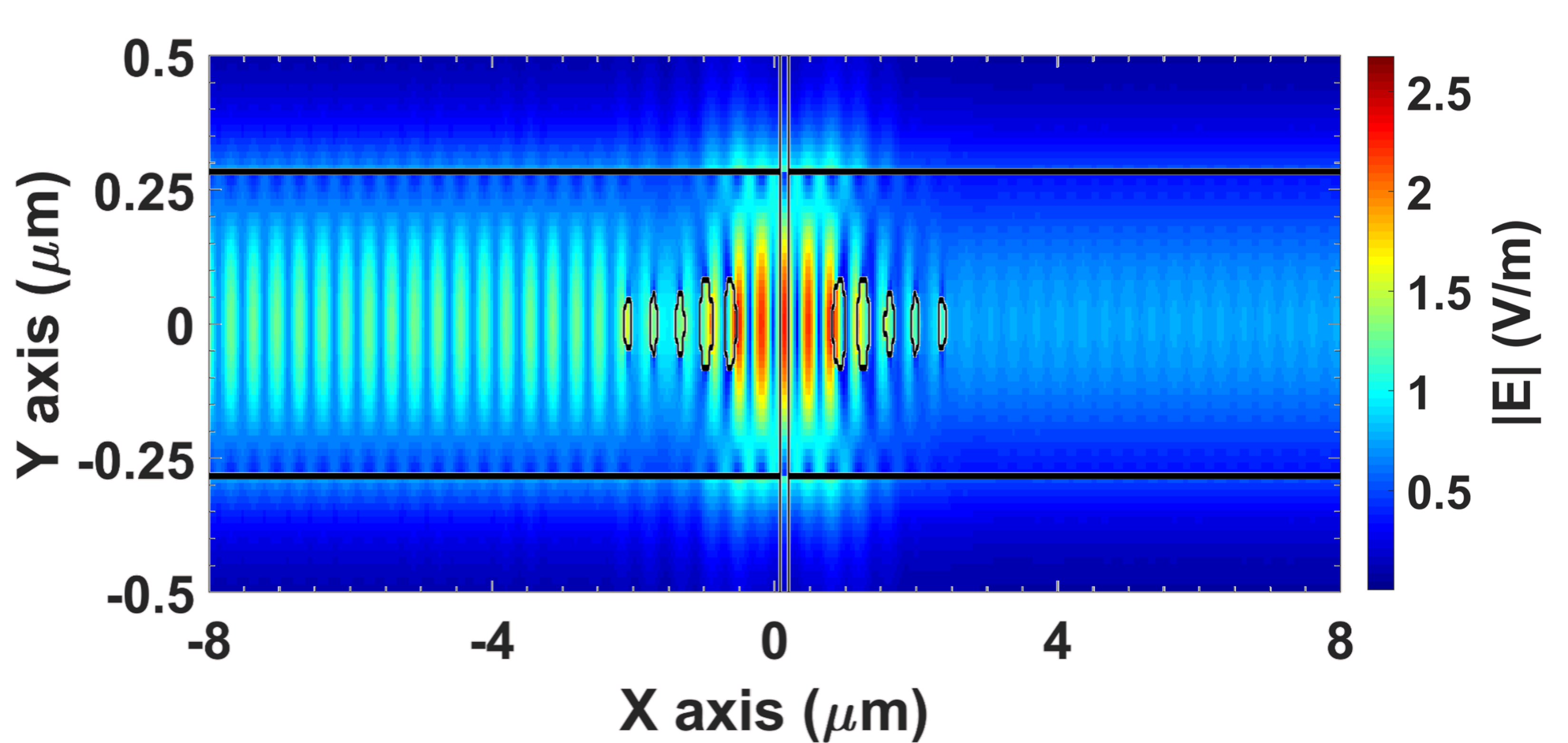}
                \caption{}
                 \label{fig 3c}
                \end{subfigure}
                 \centering
                \caption{\small Design of the photonic crystal (PhC) cavity. (a) Schematic of the PhC cavity, where the diameters are $d$\textsubscript{1-4} = 100, 110, 120, and 170 nm and the pitch, $p$, is 370 nm. The waveguide has a width and a height of $w_{1}$ = 570 nm and $h$ = 280 nm, respectively. (b) The resonance spectrum of the PhC cavity, probed in the middle of the waveguide and at the center of the cavity with a full width at half maximum (FWHM) of 44 nm. (c) The two dimensional (2D) electric field distributions in the \textbf{XY} plane through the waveguide.}
                \label{fig 3}
                \end{figure*}

\section*{SUPPLEMENTARY NOTE 3: PMOKE consideration in designing plasmonic nanoantenna} \label{s-nanoantenna}               

\begin{figure*}[h!]
             \centering
                \begin{subfigure}[h!]{\textwidth}
                \centering
                 \includegraphics[scale = 0.33]{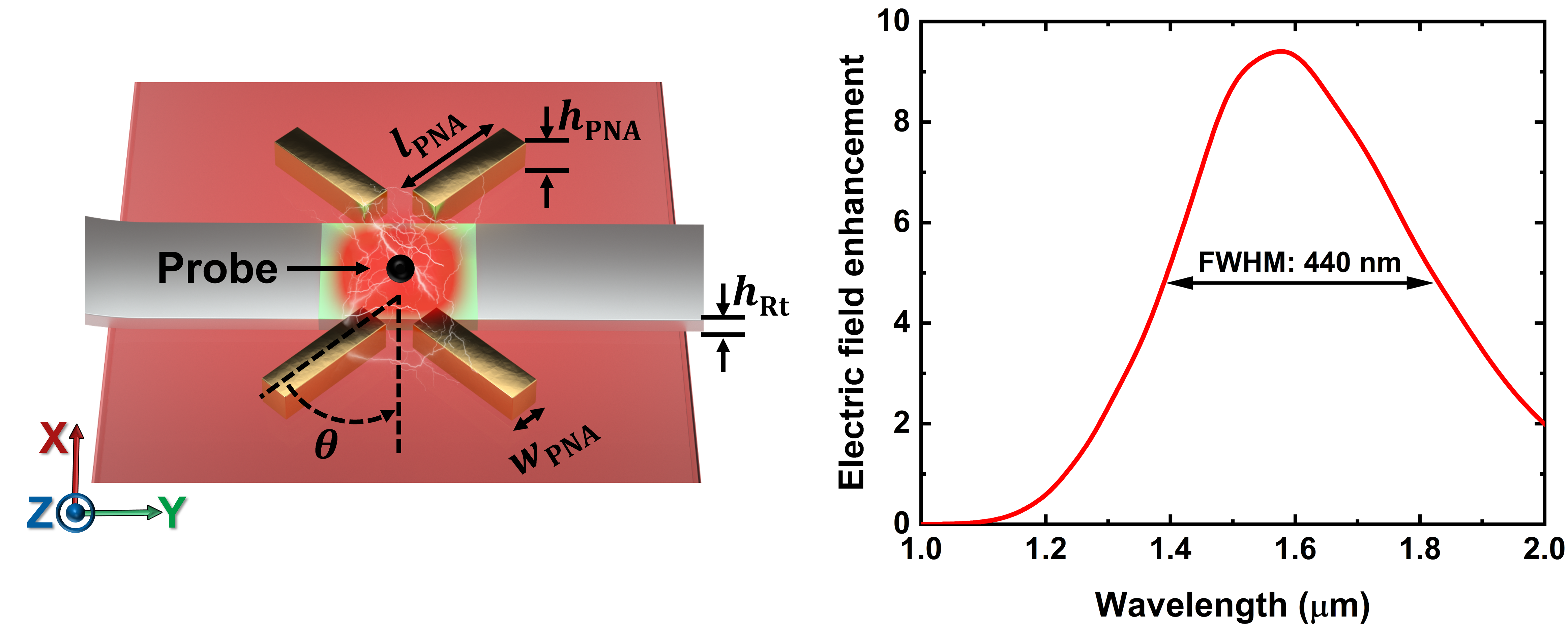}
                 \caption{}
                   \label{fig 2a}
                 \end{subfigure}
                 \begin{subfigure}[t]{0.32\textwidth}
                \centering
                \includegraphics[scale = 0.25]{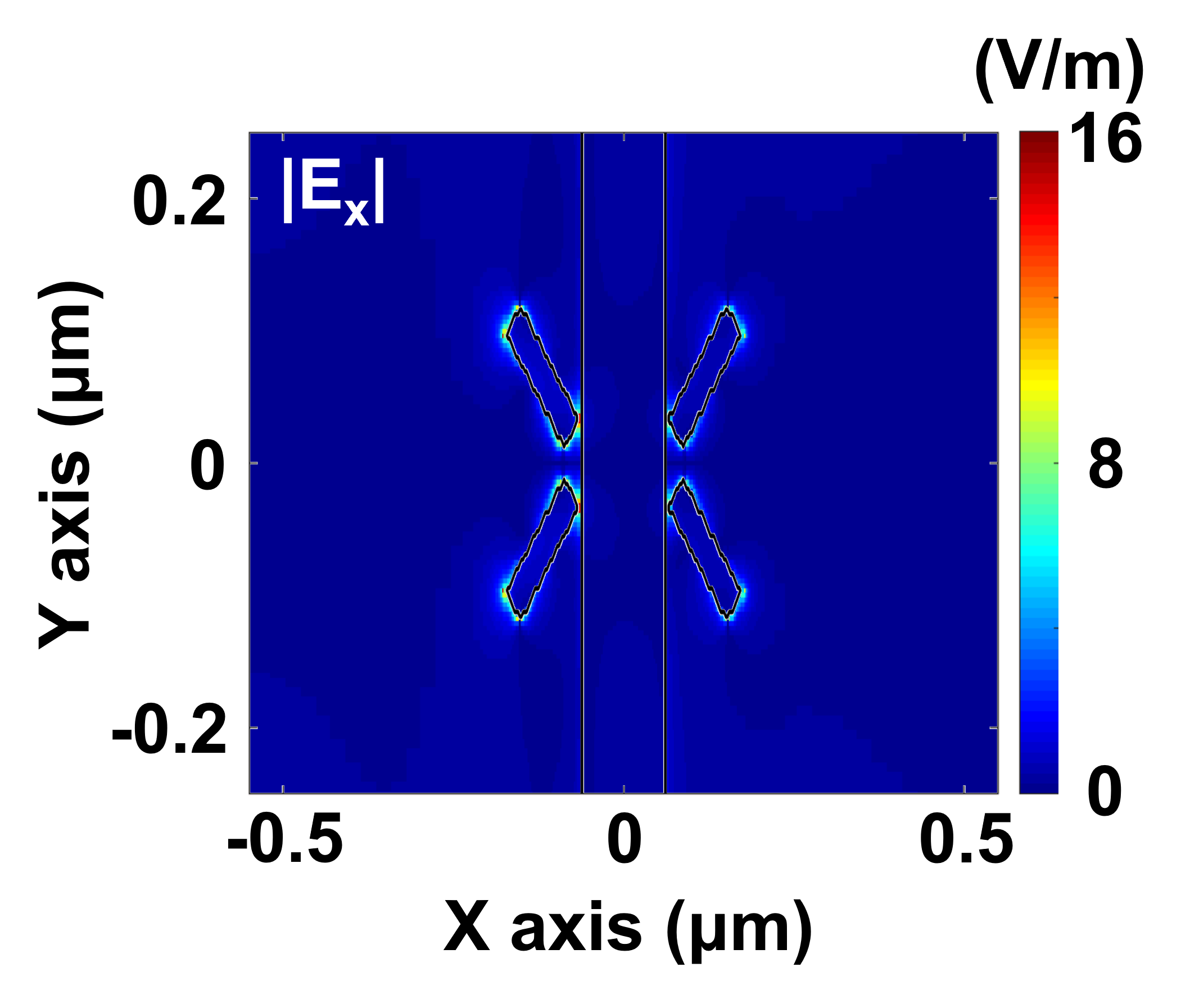}
                \caption{\hspace*{-0.6cm}}
                 \label{fig 2b}
                \end{subfigure}
                 \begin{subfigure}[t]{0.32\textwidth}
                \centering
                \includegraphics[scale = 0.25]{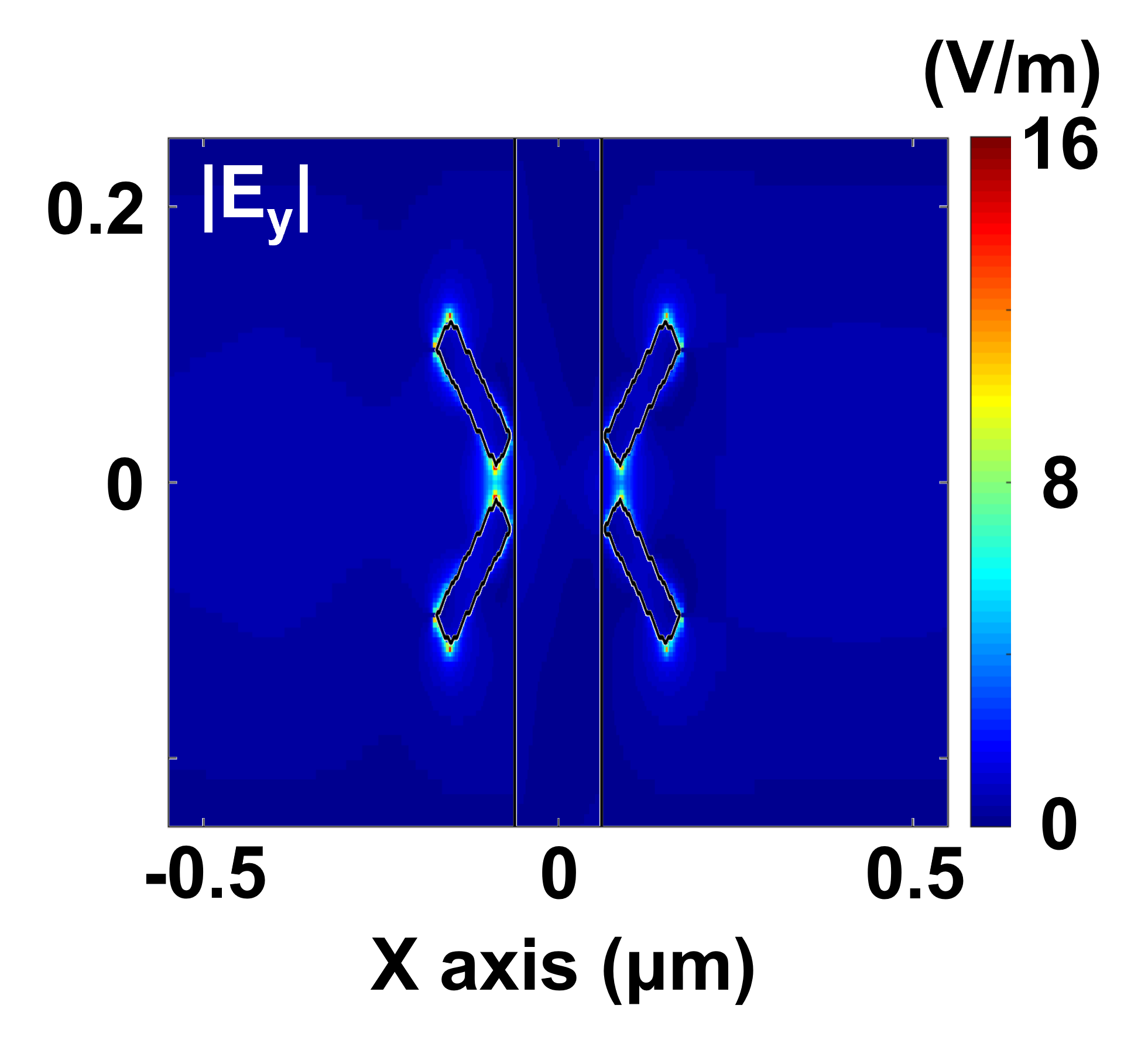}
                \caption{}
                 \label{fig 2c}
                \end{subfigure}
                 \begin{subfigure}[t]{0.32\textwidth}
                \centering
                \includegraphics[scale = 0.25]{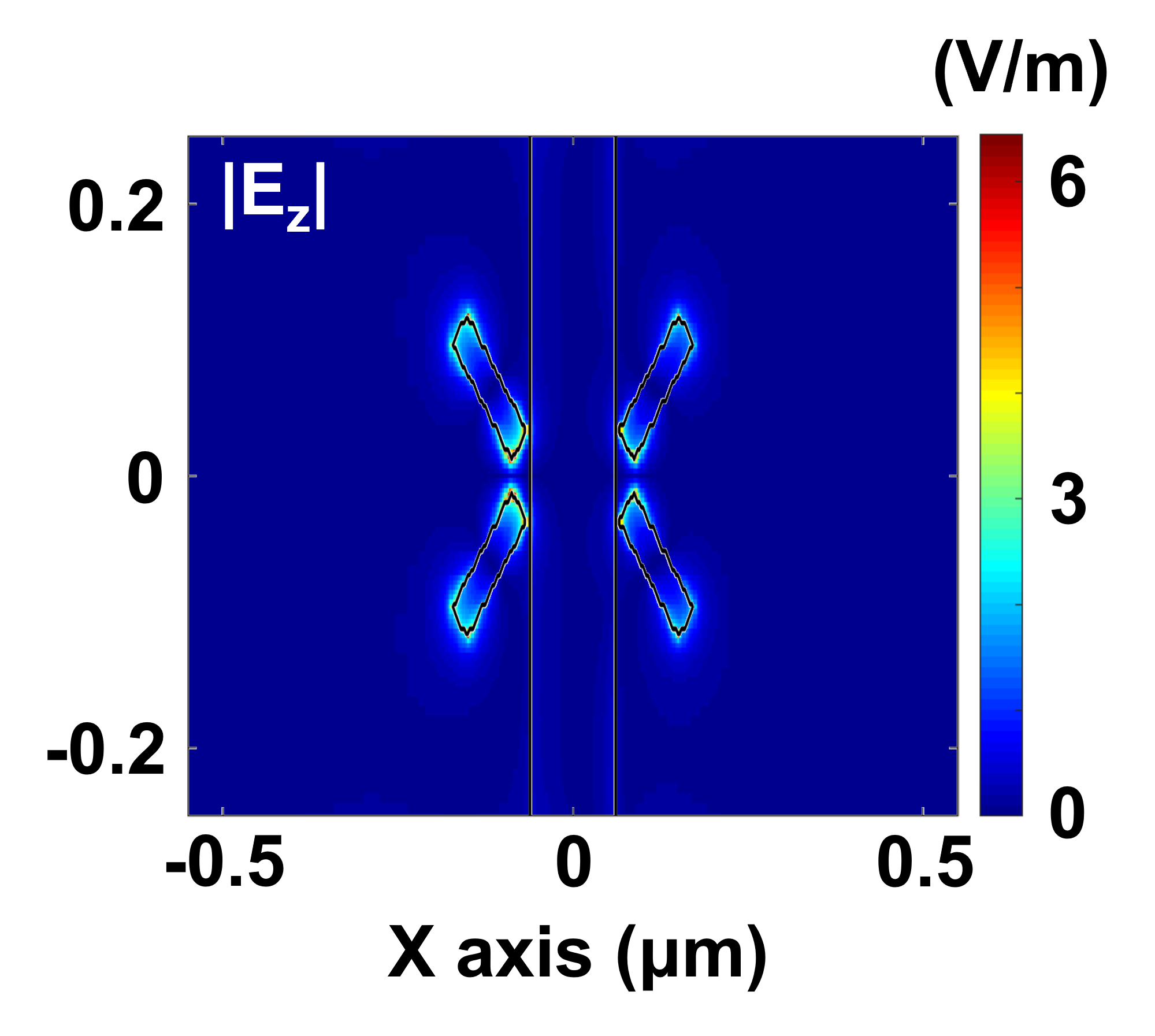}
                \caption{}
                 \label{fig 2d}
                \end{subfigure}
                \centering
                \caption{\small PMOKE consideration in designing the plasmonic nanoantenna (PNA). (a) The schematic on the left side and the resonance spectrum of PNA, probed at the center of its gap, on the right side with a FWHM of 440 nm. (b-d) The 2D electric field distribution for an input TE\textsubscript{0} mode in the \textbf{XY} plane on the top surface of the waveguide at $\lambda_0$ = 1.55 $\mu$m for \textbf{x}, \textbf{y}, and \textbf{z} components of the electric field, respectively. The optimized parameters are $l$\textsubscript{PNA} = 120 nm, $h$\textsubscript{PNA} = $w$\textsubscript{PNA} = 30 nm, $\theta$ = 45$^\circ$, and the racetrack height, $h$\textsubscript{Rt}, is 10 nm.}
                \label{fig 2}
                \end{figure*}

The permittivity tensor of the ferrimagnetic bits with PMA that we used in our modelling is defined as:
            
            \begin{equation}
                \overleftrightarrow \epsilon = 
                \begin{pmatrix}
                    \epsilon_{xx} & \epsilon_{xy} & 0\\
                    -\epsilon_{xy} & \epsilon_{xx} & 0\\
                    0 & 0 & \epsilon_{xx}
                \end{pmatrix},
                \label{Eq-2}
            \end{equation}

\noindent where the off-diagonal elements $\pm \epsilon_{xy}$ are responsible for the PMOKE. These two elements are much smaller than the diagonal elements, which make the PMOKE intrinsically weak. However, because of the localized surface plasmon resonance, the proposed plasmonic nanoantenna (PNA) with its broad resonance spectrum with FWHM of 440 nm (see the right panel in Fig. \ref{fig 2a}) can enhance and manipulate the effective polarizability of the coupled magnetic racetrack along the directions of the off-diagonal elements of the PMOKE permittivity tensor, i.e. along the \textbf{X} and \textbf{Y} axes (see Eq. \ref{Eq-2}). To do so, we configured the PNA in a double V-shaped configuration (see the left panel in Fig. \ref{fig 2a}), where the PNA elements are oriented at an angle of $\theta$ = 45$^\circ$ with reference to the waveguide direction, and the optimized length, height, and width are found to be as $l$\textsubscript{PNA} = 120 nm, and $h$\textsubscript{PNA} = $w$\textsubscript{PNA} = 30 nm. As a result, for a TE\textsubscript{0} waveguide mode with its electric field oscillating along the \textbf{Y} axis, the induced effective polarizability by the PNA becomes larger along the \textbf{X} axis than along the \textbf{Z} axis, which is essential to enhance the PMOKE in the waveguide configuration. Figure \ref{fig 2b} to \ref{fig 2d} show the electric field distribution for \textbf{x}, \textbf{y}, and \textbf{z} components of the electric field in the \textbf{XY} plane, by which one can see that the magnitude of the \textbf{x} and \textbf{y} components of the electric field, i.e. $\lvert E_x \rvert$ and $\lvert E_y \rvert$ are almost the same. In contrast, the ratio of the x component to z component is greater than 2, i.e. $\lvert E_x \rvert / \lvert E_z \rvert$ $>$ 2. So, in this case, the proposed PNA can enhance the effective polarizability of the racetrack along \textbf{X} and \textbf{Y} directions rather than the \textbf{Y} and \textbf{Z} directions. \par

\end{document}